\documentclass[12pt]{article}
\pdfoutput=1

\usepackage{epsfig,a4wide,amsmath,amssymb,amsfonts,mathrsfs,amsthm}
\usepackage{mathtools}
\usepackage{xcolor,yfonts}
\usepackage{graphicx}
\usepackage{subcaption}
\usepackage[utf8]{inputenc}
\usepackage{siunitx}
\usepackage{rotating}
\usepackage{hyperref}
\usepackage[normalem]{ulem}
\usepackage{cite}
\usepackage{float}

\newcommand{\be}{\begin{equation}}
\newcommand{\ee}{\end{equation}}
\numberwithin{equation}{section}
\newcommand{\mytitlefont}{\fontseries{mx}\selectfont}
\DeclareMathAlphabet{\titlemath}{OT1}{cmr}{mx}{n}
\newtheorem*{conjecture}{Conjecture}

\newcommand{\ignore}[1]{}

\def\l{\lambda}
\def\L{\Lambda}

\def\MA{\mathcal{A}}
\def\MB{\mathcal{B}}
\def\MC{\mathcal{C}}

\begin{document}

\begin{titlepage}

\begin{center}

~\\[2cm]

{\fontsize{20pt}{0pt} \mytitlefont  An infinity of black holes}

~\\[0.5cm]

{\fontsize{14pt}{0pt} Gary~T.~Horowitz,  Diandian~Wang and Xiaohua~Ye}

~\\[0.1cm]

\it{   Department of Physics, University of California, Santa Barbara, CA 93106}

~\\[0.05cm]

\end{center}

  \vspace{60pt}

\noindent
In general relativity (without matter), there is typically a one parameter family of static, maximally symmetric black hole solutions labelled by their mass. We show that there are situations with many more black holes. We study asymptotically
anti-de Sitter solutions in six and seven dimensions having a conformal boundary which is a product of spheres cross time. We show that the number of families of static, maximally symmetric black holes depends on the ratio, $\lambda$, of the radii of the boundary spheres. As  $\lambda$ approaches a critical value, $\lambda_{c}$, the number of such families becomes infinite. In each family, we can take the size of the black hole to zero, obtaining an infinite number of static, maximally symmetric non-black hole solutions. We discuss several applications of these results, including Hawking-Page phase transitions and the phase diagram of dual field theories on a product of spheres, new positive energy conjectures, and more.

\vfill

    \noindent

  \end{titlepage}

   \newpage

\tableofcontents
\baselineskip=16pt

\section{Introduction}

In four dimensions, the black hole uniqueness theorem \cite{Israel:1967wq,Chrusciel:2012jk} shows that in the absence of matter, the only static, asymptotically flat black hole is described by the Schwarzschild solution.
With a negative cosmological constant, black hole solutions depend on a choice of conformal metric on the boundary at infinity. If one chooses this metric to be conformal to a round $S^2$ cross time, the only known static black hole is given by the Schwarzschild-AdS solution. While it has not yet been proven to be the unique static black hole, it is certainly the only static \textit{and spherical} black hole with these boundary conditions. Both of these are one parameter families of solutions labelled by the total mass.

In higher dimensions, it is known that black holes are less unique. In this paper we describe an extreme form of this non-uniqueness. We show that there are asymptotically AdS boundary conditions such that there are an infinite number of families of black holes.  This happens when the boundary metric is conformal to time cross a product of spheres: $R \times S^m \times S^n$. Curiously, the total spatial dimension must be less than nine, $m+n < 9$ (and $m,n \ge 2)$, for this to occur. The solutions are all static and have maximal symmetry in the following sense. We assume the spheres are round, so bulk solutions with maximal symmetry will have isometry group $R \times SO(m+1) \times SO(n+1) $ and depend only on a radial coordinate. We show that the number of families of static, maximally symmetric black holes depends on the ratio, $\lambda$, of the radii of the boundary spheres. There is a critical ratio, $\lambda_{c}$, determined by the condition that the product metric on $S^m \times S^n$ is an Einstein metric. As $\l \to \lambda_{c}$ the number of families of black holes grows without bound, and at $\l = \l_{c}$ there are an infinite number of such families. Most of these families only contain small black holes. For each $\l$, there is a single family of solutions that extends to arbitrarily large black holes.

In each family of black holes, one can take the size of the black hole  to zero, and thus obtain an infinite number of static, non-black hole solutions as well.  Since we have a single asymptotic boundary, in the non-black hole solutions one of the spheres must shrink to a point in the interior. When one sphere on the boundary is much larger than the other (i.e., $\l$ is either very large or very small) the smaller sphere is the one that shrinks. However, when they are comparable in size ($\l \sim \l_{c}$) either sphere can shrink, and it turns out that they can do so in many inequivalent ways. When one sphere shrinks to a point, the other sphere becomes a minimal surface at that point. The solutions are distinguished by the size of that minimal surface.

An important precursor to this work is the paper by Aharony et al.~\cite{Aharony:2019vgs}, who studied Euclidean, asymptotically AdS solutions with conformal boundary $S^m \times S^n$ (without a time direction). They found that when $m+n < 9$ (and $m,n \ge 2$) the number of solutions increases as $\l \to \l_{c}$, and becomes infinite at $\l = \l_{c}$. We first add a time direction (in both the bulk and boundary) and find similar behavior for the non-black hole spacetimes. We then show that one can add black holes to each of these background solutions. These black holes have horizon topology $S^m \times S^n$, and when their area goes to zero, one sphere on the horizon becomes much smaller than the other. In this regime one expects the black hole to be unstable due to a Gregory-Laflamme instability \cite{Gregory:1993vy}. The stable solution should be a small black hole with $S^{m+n}$ topology. This is yet another class of black holes with the same boundary conditions but less symmetry, which we will not discuss here.  

\begin{figure}
    \centering
    \includegraphics[width=0.65\textwidth]{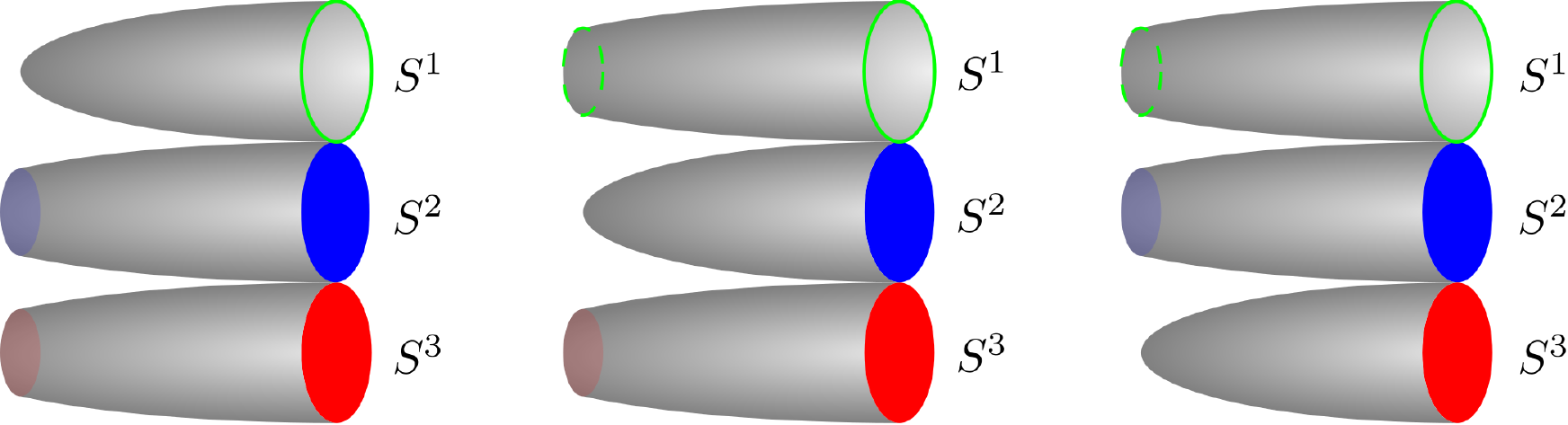}
    \caption{The topology of the various bulk solutions we will construct: The left column shows the topology for a (Euclidean) black hole solution where the thermal $S^1$ is contractible. The middle column illustrates one type of non-black hole solution where the $S^2$ is contractible and hence has topology $S^1\times R^3\times S^3$, and the right column shows the other non-black hole solution where the $S^3$ is contractible and hence has topology $S^1\times S^2\times R^4$.}
    \label{fig:cigars}
\end{figure}

We will study in detail two cases: $S^2\times S^3$ and $S^2\times S^2$. Since the analysis and results are  similar, we describe $S^2\times S^3$ in the main text, and $S^2\times S^2$ in an appendix. There is a simple analytic black hole solution when $\l = \l_c$, which becomes singular when the black hole shrinks to zero size. We discuss this solution first in Sec.~\ref{sec:analytic}.  We then  construct several families of black holes and non-black hole solutions numerically for various $\l$ in Sec.~\ref{sec:zero_temp} and \ref{sec:BH}, and show that the number of such families diverges as $\l \to \l_c$. To construct the black holes, it is convenient to analytically continue  (and periodically identify) time, so the boundary becomes $S^1\times S^2\times S^3$. The black hole solutions are now simply the ones where the $S^1$ pinches off in the interior before either sphere. Fig.~\ref{fig:cigars} illustrates the three classes of solutions we will study. 

There are many applications of these results which we  discuss in Sec.~\ref{sec:app}. First, we compute the energy of the non-black hole solutions. Since the minimum energy solution with fixed boundary conditions should be a smooth, static,  solution with maximal possible symmetry \cite{Sudarsky:1992ty, Hertog:2005hm}, our solution with lowest energy should be the ground state. This leads to new positive energy conjectures for asymptotically AdS gravity with boundary $R\times S^2 \times S^3 $.  We will find that the transition between one sphere being contractible in the ground state and the other sphere being contractible occurs at a value $\l = \l_t$, which is slightly larger than $\l_c$. In particular,  there are two solutions  with the same minimum energy when $\l = \l_t$. Thus, unlike most positive energy conjectures, in this case the minimum energy solution is not unique.

As a second application, we study the implications for a six dimensional dual field theory on $R\times S^2 \times S^3 $. The minimum energy solutions are dual to the vacuum state of this theory, and their energy can be viewed as a Casimir energy of the field theory. We will see that this Casimir energy is positive for very large or very small $\l$ and is negative for $\l \sim \l_c$. The fact that the bulk geometry changes  abruptly at $\l_t$ means that there is a quantum phase transition at this point. We also compute the Euclidean action of the black hole solutions and determine the solutions that dominate a canonical ensemble at fixed temperature $T$. At large $T$, the (unique) large black hole with the given $\l$ dominates, but as one lowers $T$, there is a Hawking-Page phase transition to a gas of gravitons on the ground state. The numerous small black holes never dominate the canonical ensemble. Note that there is a one parameter family of Hawking-Page phase transitions labelled by $\l$. 

We also study the microcanonical ensemble and determine which small black holes have the greatest entropy for given energy. This is not reliable for very small $E$ since we study $S^2\times S^3$ black holes and we expect $S^5$ black holes will have larger entropy. However, we find an interesting ``phase transition" at intermediate $E$ where two different families of $S^2\times S^3$ black holes exchange dominance. Finally, we use these solutions to learn something about confining gauge theories on de Sitter space, and limitations of the state operator correspondence for conformal field theories.

Before proceeding, we comment on  related work. The simplest example where the ground state is not AdS, is when the boundary contains a spatial circle such as $R^{n} \times S^1$ or $R \times T^n$. In this case the ground state is the AdS soliton \cite{Horowitz:1998ha} and has negative energy. At finite temperature there is a phase transition between a planar black hole and thermal gravitons on the AdS soliton. A generalization of the AdS soliton where the contractible circle is replaced by a contractible sphere (so the boundary is $R^n\times S^m$) was constructed in \cite{Kleihaus:2010am}, along with black holes with the same conformal boundary (see also \cite{Harlow:2018tng}). The case where the boundary topology is $R\times S^1\times S^2$ was studied in \cite{Copsey:2006br}. A similar case with the $S^2$ factor replaced by $S^3$ was studied in \cite{Hickling:2016mzp}. Euclidean solutions with boundary $S^m \times S^n$  (without a time direction) were studied in \cite{Aharony:2019vgs} (see also \cite{Blackman:2011in} for when $m=2$ and \cite{Kiritsis:2020bds} for when $m=n=2$). Finally, when the boundary is $R\times S^2\times S^2$, which is the case we study in App.~\ref{sec:S2S2}, black hole solutions were discussed in \cite{Hickling:2016mzp}, but the dependence of the solutions on the ratio of the spheres at the conformal boundary was not analyzed. For special ratios of the boundary spheres, Euclidean analytic solutions and their analytic continuation to Lorentzian spacetimes were discussed in \cite{Dibitetto:2020csn}.

\section{Analytic solutions}\label{sec:analytic}

We are interested in vacuum solutions to general relativity with a negative cosmological constant that have a conformal boundary which is a product of spheres cross time.
Although most of the solutions we study will be numerical, there is a simple analytic family of solutions with these boundary conditions which we describe first. Since these solutions exist in all dimensions greater than five,  we begin by considering general $D = d+1$ dimensions.

We start with the usual $D$-dimensional Schwarzschild-AdS solution:
\begin{equation}
\label{eq:SchAdS}
    ds^2 = - f(r) dt^2 +\frac{dr^2}{f(r)} + r^2 d\Omega_{d-1}^2,
    \end{equation}
where
\begin{equation}
\label{eq:SchAdSf}
    f(r) = r^2 +1 - \frac{r_0^{d} + r_0^{d-2}}{r^{d-2}},
\end{equation}
and we have labelled the solutions by the horizon radius $r_0$, and set the AdS radius of curvature to one. The key observation is that this metric remains a solution if we replace the unit $(d-1)$-sphere with any other Einstein metric with the same scalar curvature{, $\mathcal{R}$}. Given two (round)  spheres $S^m$ and $S^n$ with radii $r_m$ and $r_n$, and $m+n = d-1$, their product $S^m\times S^n$ will satisfy ${\cal R}_{ij} \propto g_{ij} $ if 
\begin{equation}
\label{eq:critratio}
    \frac{r_m}{r_n} = \sqrt \frac{m-1}{n-1}.
\end{equation}
Requiring that the scalar curvature agree with the unit $(d-1)$-sphere, i.e., ${\cal R} = (d-1)(d-2)$, fixes the radii, so we get the solution
\begin{equation}
\label{eq:AdSpq}
    ds^2 = - f(r) dt^2 +\frac{dr^2}{f(r)} + r^2 \left[ \frac{m-1}{d-2}\ d\Omega_m^2  + \frac{n-1}{d-2} \ d\Omega_n^2 \right].
    \end{equation}
The stability of this black hole has been studied \cite{Gibbons:2002pq,Hartnoll:2003as} and it was shown that it is stable for large $r_0$ but becomes unstable for small $r_0$ (provided $m+n<9$).
Setting $r_0 = 0 $ in $f(r) $ yields a solution without a black hole, but it has a (naked) curvature singularity at $r= 0$.
    
We are mainly interested in  $D = 7$ with boundary $R\times S^2 \times S^3$. In this case, the critical ratio of radii where the product of spheres has an Einstein metric is given by $(r_2/r_3)^2 = 1/2$. We will see that \eqref{eq:AdSpq}  is not the only family of black holes with this ratio of spheres on the conformal boundary. In  fact, we will argue that  there are an infinite number of other solutions. Similarly, we will construct many solutions without black holes with this asymptotic ratio, that have no naked singularities. Again, there are  an infinite number of them.

Since we want to consider all ratios of sphere radii, it will be convenient to choose a different radial coordinate than the one above.  We will   often choose $r$ so that the metric on $S^3$ is just $r^2 d \Omega_3^2$.   Compared to \eqref{eq:AdSpq} this involves rescaling the radial coordinate (and the time coordinate), so the constant in $f(r)$ is no longer one. Instead we have
\begin{equation}
\label{eq:ourform}
    ds^2 = - \mathcal{A}(r) dt^2 +\frac{dr^2}{\mathcal{A}(r)} + r^2 \left [\frac{1}{2}\ d\Omega_2^2   + d\Omega_3^2\right],
    \end{equation}
where 
\begin{equation}
\label{eq:SchAdSA}
    \mathcal{A}(r) = r^2 +\frac{1}{2}  - \frac{2r_0^6 + r_0^4}{2r^{4}}.
\end{equation}
This black hole has a temperature
\begin{equation}
\label{eq:analT}
   T = \frac{1+3r_0^2}{2\pi r_0}.
\end{equation}
To compute the Euclidean action for this (as well as our other) asymptotically AdS metrics, we  perform holographic renormalization by adding boundary counter terms to the usual Einstein-Hilbert action with Gibbons-Hawking-York boundary term. We explain how to  do this in App.~\ref{sec:RG}.
The result is 
\begin{equation}
\label{eq:analact}
   I = -\frac{\pi^3r_0\left(5 - 32r_0^4 + 64 r_0^6\right)}{128\left(1+3r_0^2\right)} ,
\end{equation}
where we have set $G=1$.

\section{Non-black hole solutions}\label{sec:zero_temp}

Consider a conformal boundary $R\times S^2\times S^3$ with  representative metric
\begin{equation}
\label{eq:g_bdy_S2S3}
    ds^2|_{\partial \mathcal{M}}= - dt^2 + \l^2 d\Omega_2 + d\Omega_3.
\end{equation}
This describes a static product of the two spheres, each with its own spherical symmetry. We have used conformal rescaling to make the three-sphere of unit size. The only free parameter for the family of metrics is $\l$, the ratio between the two sphere radii. 

We begin by finding all candidate bulk ground states in Lorentzian signature for given $\l$'s. With this boundary condition, the bulk ground state is expected to have maximal symmetry, by which we mean both time translation symmetry and $SO(3)\times SO(4)$. Now, there are two possible ways for the bulk to have a static and smooth (non-black hole) geometry: either the $S^2$ or $S^3$ must smoothly pinch off in the bulk, i.e., the size of one of the spheres must shrink to zero. 

A maximally symmetric bulk metric with contractible $S^2$ can be written as
\begin{equation}
\begin{aligned}\label{AdS7_S2metric}
    ds^2&= -\mathcal{A}(r)dt^2+\frac{dr^2}{\mathcal{A}(r)\mathcal{B}(r)}+\mathcal{C}(r)\l^2 d\Omega_2+r^2 d\Omega_3, \\
    \mathcal{A}(r)&=r^2+\sum_{n=-1}^\infty \frac{a_{n}}{r^n},
    \quad\mathcal{B}(r)=1+\sum_{n=1}^\infty \frac{b_{n}}{r^n},\quad
    \mathcal{C}(r)=r^2 +\sum_{n=-1}^\infty \frac{c_{n}}{r^n},
\end{aligned}
\end{equation}
where we have specified the leading order coefficients for each of the functions to ensure the correct boundary metric \eqref{eq:g_bdy_S2S3}.  The choice of the radial coordinate $r$ favors $S^3$ over $S^2$ because it gives the proper size of the three-sphere at any radius (in particular, $S^3$ shrinks to zero size at $r=0$). With this choice, for the solution to have contractible $S^2$, the two-sphere needs to shrink to zero at some $r_0>0$. In other words, the function $\mathcal{C}(r)$ should have the property that $\mathcal{C}(r_0)=0$ for some $r_0>0$ (whereas $\mathcal{A}(r)$ and $\mathcal{B}(r)$ should be positive for all $r\ge r_0$). The spacetime only exists for $r \ge r_0$.

\begin{figure}
    \centering
    \includegraphics[width=9cm]{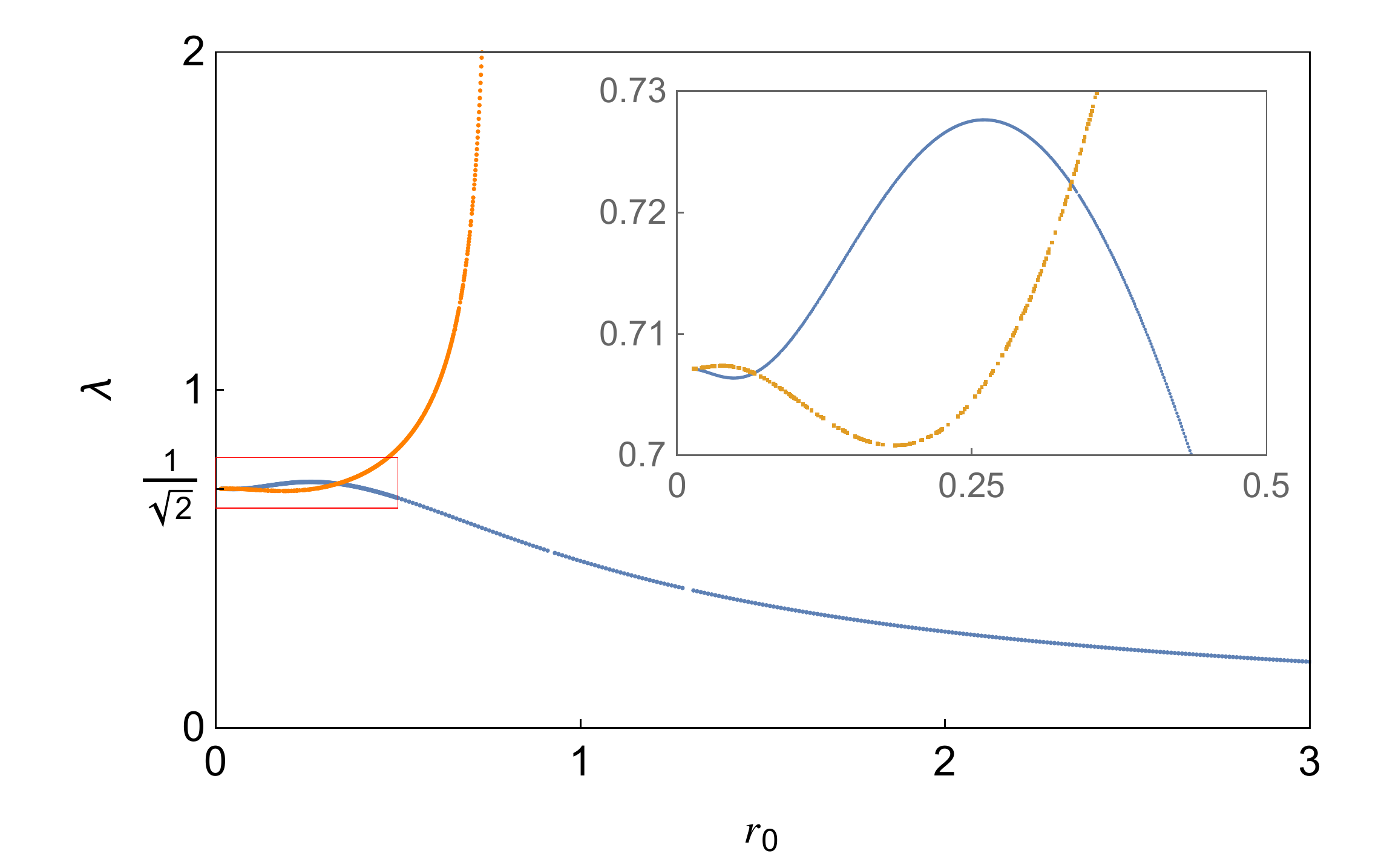}
    \caption{The space of static, non-black hole solutions. The blue curve corresponds to $S^2$-contractible solutions, and the orange curve represents $S^3$-contractible solutions. There are an infinite number of oscillations for small $r_0$ and near $\l \approx\sqrt{0.5}$. The insert shows the first two.}
    \label{fig:AdS7_oscillation}
\end{figure}

In App.~\ref{sec:num_ODE}, we describe our numerical algorithm for solving the Einstein equations as a set of decoupled ODEs. It turns out convenient to use $r_0$ to label all the solutions. For each $r_0\in \mathbb{R}_+$, there is a unique choice of $\l$ such that a smooth solution (with this topology) exists. We can think of all such solutions as lying on a curve in the $(\l,r_0)$ plane. Fig.~\ref{fig:AdS7_oscillation} displays this curve (blue), computed numerically.
As the figure demonstrates, there is exactly one solution for each $r_0$. 

Although $r_0$ is a good parameter to label all the solutions, it is not a variable on the boundary (it depends on the bulk coordinate system). Therefore, if we want to ask what bulk solutions correspond to a given boundary geometry, $\l$ should be regarded as the independent parameter. From this viewpoint, Fig.~\ref{fig:AdS7_oscillation} tells us something quite interesting. For small values of $\l$, there is only one solution with $S^2$ contractable; for large values of $\l$, no such solution exists; however, when $\l$ is close to a special value $\l_c=\sqrt{0.5}$, there can be more than one solution for each $\l$. In fact, due to the oscillatory shape of the curve, the number of solutions grows as $\l$ approaches this special value. Interestingly, infinitely many oscillations were found for Euclidean boundaries with topology $S^{m}\times S^{n}$ ($m,n \ge 2$, $m+n<9$) \cite{Aharony:2019vgs}. This figure suggests that this also happens with the extra time direction that we have included. We provide an argument for this in App.~\ref{sec:pert}. As a result of this infinite oscillation, the number of maximally symmetric solutions for a given $\l$ grows without bound as $\l\to \l_c$. 

We can also obtain $S^3$-contractible solutions similarly. In this case, it is more convenient to use the following ansatz:
\begin{equation}\label{AdS7_S3metric}
    ds^2 = -\mathcal{A}(r)dt^2+\frac{dr^2}{\mathcal{A}(r)\mathcal{B}(r)}+ r^2 \l^2 d\Omega_2 + \mathcal{C}(r) d\Omega_3.
\end{equation}
Our radial coordinate $r$ now favors $S^2$, and solutions with contractible $S^3$ now have $r_0>0$ where $r_0$ is now related to the size of the two-sphere when $S^3$ shrinks to zero. As a result, an $S^2$-contractible solution and an $S^3$-contractible solution with the same $r_0$ need have no relation to each other. Nevertheless, we will abuse the notation, and again use $r_0\in\{0,\infty\}$ to label all the solutions. The numerically computed curve in the  $(\l,r_0)$ plane is shown in Fig.~\ref{fig:AdS7_oscillation} (orange), superposed onto the earlier plot for $S^2$-contractible solutions.

Similar to the case studied in \cite{Aharony:2019vgs}, in the strict $r_0\to0$ limit, both curves approach $\l\to\l_c$. In this limit, since $r_0$ is the size of one of the spheres as the other sphere shrinks to zero size, both spheres now shrink to zero size, making the curvature singular at this point. In fact, it is given by the $r_0\to 0$ limit of the analytic solutions described in Sec.~\ref{sec:analytic} by \eqref{eq:ourform} and \eqref{eq:SchAdSA}. This is a very convenient fact because, with an analytic expression for the singular solution, it is possible to study perturbations analytically, which, combined with dimensional analysis, allows us to prove the existence of an infinite number of oscillations in App.~\ref{sec:pert}.

Given that there are many static, nonsingular, bulk solutions for certain $\l$, it is natural to ask  which one has the lowest energy. We will discuss this in Sec.~\ref{ssec:pos_ene_theo}. However, before doing so, in the next section we show that each of these solutions gives rise to a one-parameter family of black hole solutions.

\section{Black hole solutions}\label{sec:BH}
We now look for black hole solutions by going to Euclidean signature. This has the added benefit of getting the temperature almost for free. For Euclidean black holes, the compactified time circle $S^1$ pinches off in the bulk. In the language of Fig.~\ref{fig:cigars}, they are the first type of solutions.

To begin with, we choose the following ansatz:
\begin{equation}\label{AdS7_BHmetric}
    ds^2 = \mathcal{A}(r)d\tau^2+\frac{dr^2}{\mathcal{A}(r)\mathcal{B}(r)}+\mathcal{C}(r)\l^2 d\Omega_2 + r^2d\Omega_3,
\end{equation}
where $\MA(r_0)=0$ for some $r_0>0$ and $\MC(r\ge r_0)>0$. We have again used the same notation, $r_0$, even though it now has a new definition: the size of the three-sphere at the black hole horizon. The solutions are obtained by solving the Einstein equation numerically as before. To be concrete, we describe a procedure of conveniently solving them in Sec.~\ref{ssec:num_BH}.

\begin{figure}
    \centering
    \includegraphics[height=6.2cm]{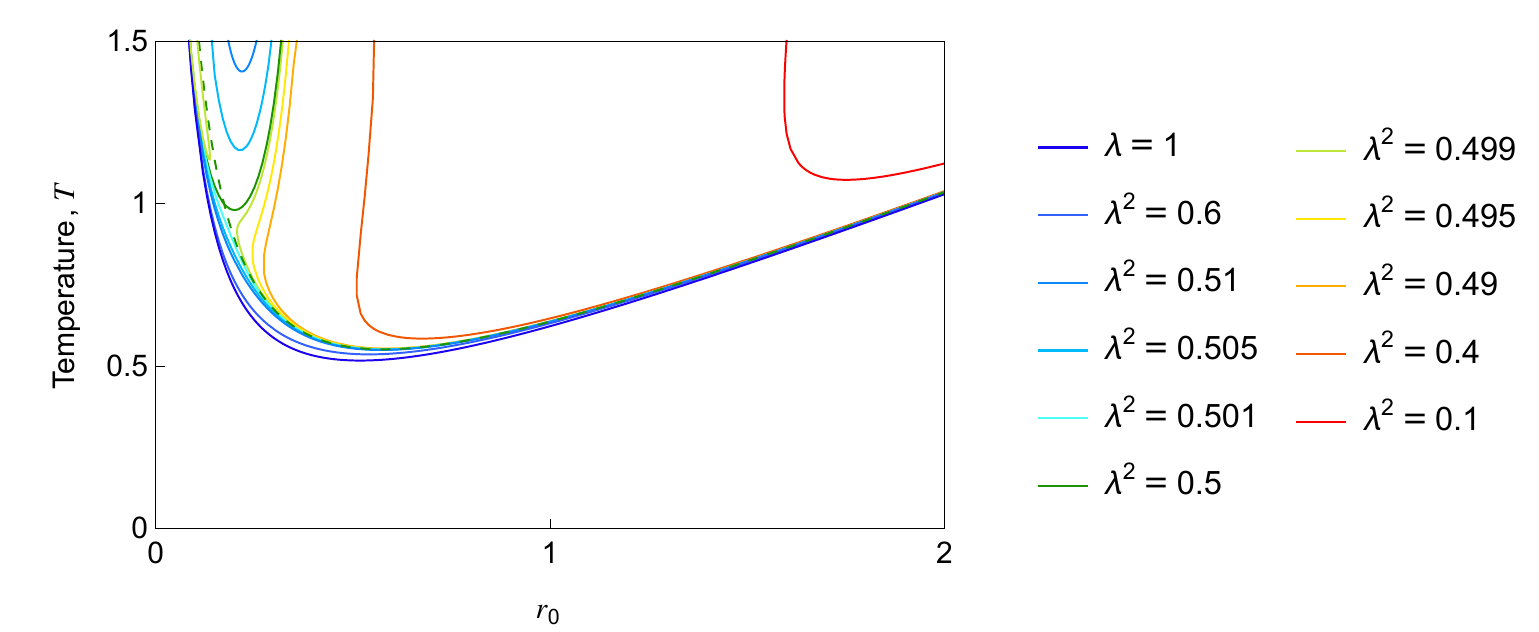}
    \caption{Temperature versus $r_0$ for various $\l$. The dashed curve corresponds to the analytic solution.}
    \label{fig:AdS7_Tvsr0}
\end{figure}

There are a two-parameter family of black hole solutions, and it is convenient to label them by $r_0$ and the temperature $T$. In Fig.~\ref{fig:AdS7_Tvsr0}, we show lines of constant $\l$ in this space. The analytic solution with $\l = \l_c$ is shown in dashed emerald green. Recall from Sec.~\ref{sec:analytic} that the two spheres for the analytic branch have fixed ratio of radii everywhere in the bulk, including in particular at the horizon. Therefore, $r_0$, the size of the three-sphere at the horizon, is a good indicator of the area of the horizon ($\sim r_0^5$). This branch starts at arbitrarily small $r_0$ where the black hole is arbitrarily hot, grows to a minimum temperature at a scale comparable to the AdS radius, and keeps growing to arbitrarily large size with again arbitrarily high temperature. Thus, the analytic branch is similar to the   familiar case of black holes with conformal boundary $S^1 \times S^{d-1}$. Indeed, for most values of $\l$, there is only a one-parameter family of solutions with similar behavior.

\begin{figure}
    \centering
    \includegraphics[height=8cm]{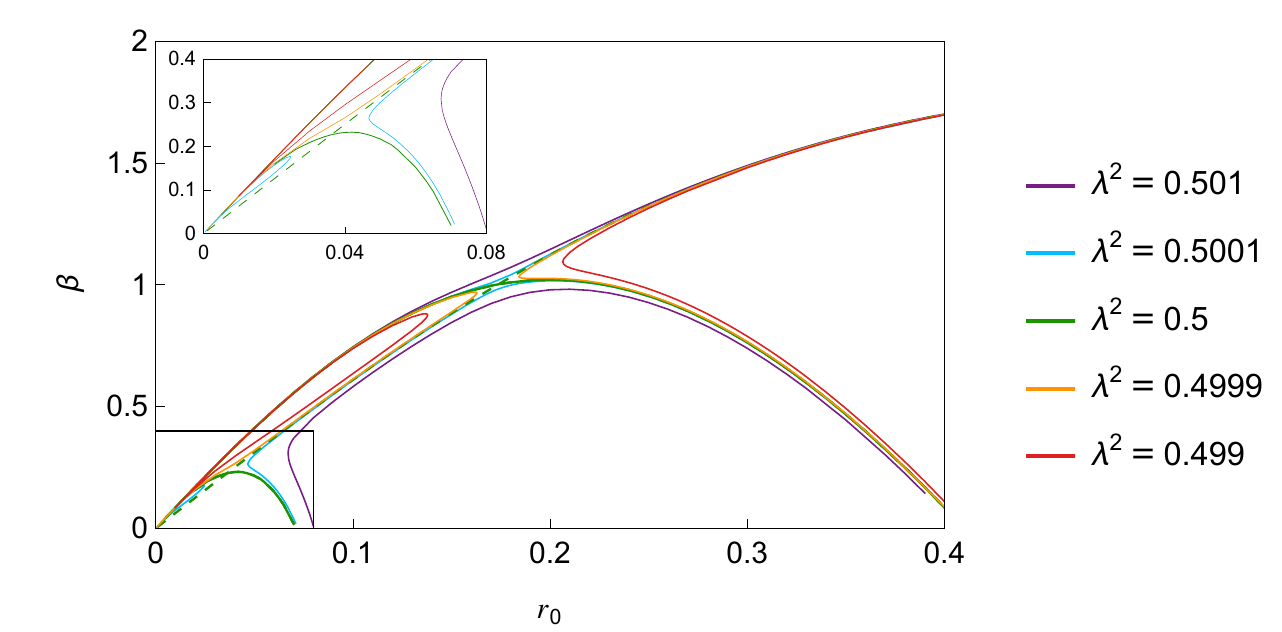}
    \caption{Inverse temperature versus $r_0$ for various $\l$. The dashed curve corresponds to the analytic solution. The plot focuses on small $r_0$ and $\l\approx \l_c$. Several branches of solutions with the same $\l$ are now visible.}
    \label{fig:AdS7_Betavsr0}
\end{figure}

\begin{figure}
    \centering
    \includegraphics[width=0.5\textwidth]{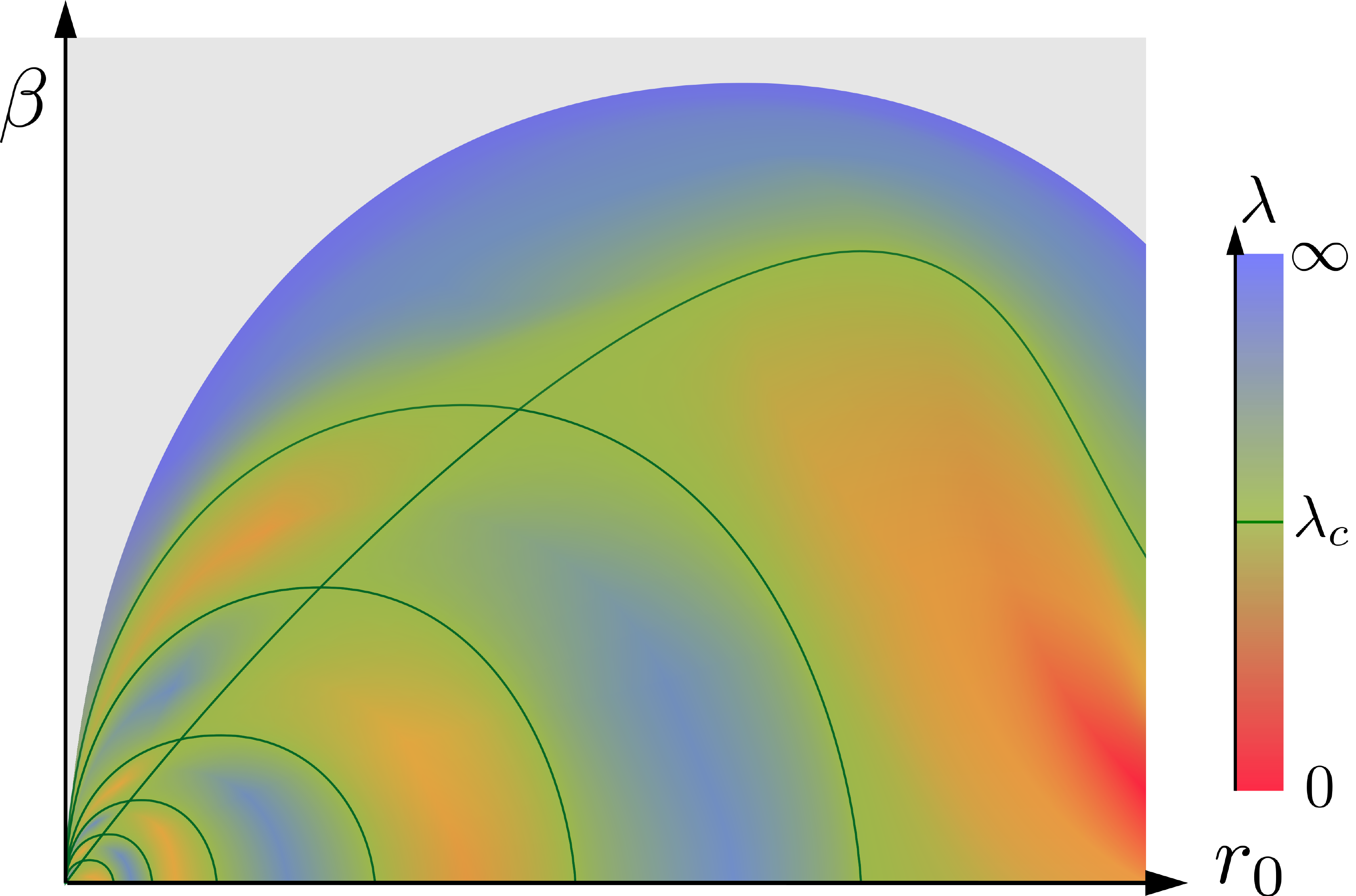}
    \caption{A sketch of  the solution space, not to scale and zoomed in on the small-$r_0$ side for visibility. The solutions for the special sphere ratio $\l=\l_c$ have been highlighted in emerald. They separate the solution space into infinitely many `cells'. The coloring correlates with $\l$. For very large (small) $\l$, there is only one solution near the upper rim (lower right corner) of the figure. For $\l\sim \l_c$, there can be more than one branch, residing in a subset of cells in a chequered fashion. Moving to the lower left corner, the color range narrows (closer to green), meaning that $\l$ needs to be very close to $\l_c$ to have solutions in those cells. There are no solutions in the grey region. }
    \label{fig:branches}
\end{figure}

However, for $\l$ close to $\l_c$, there are multiple families of solutions that show up at small $r_0$ and high $T$. This is shown in Fig.~\ref{fig:AdS7_Betavsr0}, where we have replaced $T$ with $\beta = 1/T$, and focused on the small $r_0$ region. The analytic solution with $\l = \l_c$ is again shown in dashed emerald green. But
as can be seen in the figure, there are many other branches of solutions with this same $\l$ (shown as solid emerald lines).  In fact, there are infinitely many more branches. Each of these numerical branches intersects the analytic branch exactly once (at least for the ones verified numerically). They all start at infinite temperature ($\beta = 0$) and zero $r_0$. They then grow in $r_0$ with decreasing temperature. After they reach some minimum temperature, they increase in temperature again ending at $\beta = 0$ and finite $r_0$.\footnote{\label{ft:finite_r0}It may seem strange to have black holes approach infinite $T$ at fixed $r_0$, but remember we have chosen coordinates that favor $S^3$.   As the size of the three-sphere ($r_0$) grows along such a branch, the two-sphere shrinks to zero size, so the black hole is indeed becoming small when $\beta \to 0$.} We provide a sketch for this in Fig.~\ref{fig:branches} based on extrapolation of the numerical results and an argument which we give at the end of this section. 

We have just described the solutions at the special $\l=\l_c$. Interestingly, this value is not just special due to the existence of the analytic solution, but it is the only value for which the branches of solutions intersect in the  $\beta$-$r_0$ plot. This can be seen in Fig.~\ref{fig:AdS7_Betavsr0} and is illustrated in Fig.~\ref{fig:branches}. The solutions for the special $\l_c$ separates the solution space into infinitely many subregions, or `cells'. Neighboring regions always have opposite signs of $\l-\l_c$.

Let us now be more specific about the types of cells in Fig.~\ref{fig:branches}. The first special cell is the uppermost one. It contains one branch of solutions for each value of $\l>\l_c$. Each of these branches starts at $r_0=0$ with infinite temperature (as expected for black holes with zero area) and extends all the way to infinite $r_0$. As $\l\to\infty$, the curves moves up towards the grey region but never reaches it. The second special cell is the one at the lower right corner of this figure. This contains one branch of solutions for each value of $\l<\l_c$. Each of these branches starts at some finite $r_0$ with infinite temperature (see footnote~\ref{ft:finite_r0}).  These branches also extend to infinite size. Put together, these two special cells (along with the analytic branch) contain all the branches that have arbitrarily large black holes. All other branches have a maximum horizon size, which we now describe. 

Above the analytic branch, there are infinitely many cells besides the special one. Each branch in these cells start at $(r_0=0,\beta=0)$, grows in $r_0$ to some maximum value, and comes back to $(r_0=0,\beta=0)$. Along the whole branch, the size of $S^2$ stays nonzero and finite. The two end points of each of these branches are identified with two non-black hole solutions we found in Sec.~\ref{sec:BH}, both with contractible $S^3$. Similarly, below the analytic branch, there are again infinitely many cells besides the special one. As is clear from the figure, $r_0$ remains positive and finite for all these solutions. Similar to the discussion of the numerical solutions with $\l=\l_c$ above, a feature hidden from the view is that the size of the $S^2$ on the horizon goes to zero as $\beta \to 0$ at $r_0 > 0$. The solution again reduces to one of the non-black hole solutions in these limits, both now with contractible $S^2$. Therefore, for $\l \ne \l_c$, the branches that do not reach large black holes connect two non-black hole solutions with the same topology. Contrast this with any of the numerical branches at $\l =\l_c$: each of those branches connects two
non-black hole solutions with different topology since the $S^3$ pinches off at one end and the $S^2$ pinches off at the other.

Finally, we now extend the numerical evidence that the number of families of black holes increases as $\l \to \l_c$, and argue that there are an infinite number families when $\l = \l_c$. This is simple once we have established the infinite number of non-black hole solutions, which we do in App.~\ref{sec:pert}. Basically, we want to put a small black hole in each one. Actually, as we now explain, we really put a small black brane wrapped around the noncontractible sphere.  Start with any $S^2$-contractible non-black hole solution, and let $r_0$ be the radius of the $S^3$ at that point. Then at each point on $S^3$ one can add a small four-dimensional Schwarzschild black hole with radius  $r_h \ll r_0$. This produces a large change in the metric near the horizon, but only a small change for $r \approx  r_0$, so the asymptotic ratio of spheres is very similar to the non-black hole solution. Since the curvature on the $S^3$ is small compared to the curvature near the horizon,  it looks locally like a three-dimensional black brane.  Similarly, we can do the same thing when $S^3$ pinches off, adding a small five-dimensional black hole at each point of  the minimal $S^2$. At least when adding very small black holes, different non-black hole solutions will give rise to different  black holes by following this procedure, thus generating infinitely many one-parameter families of black hole solutions. Numerically, we saw that all  of these one-parameter families (except the ones that extend to infinite size) connect at finite mass in pairs, but that at most reduces the number of families by a half, which is still infinite. Thus, we have established that an infinity of black holes do exist. 

\section{Applications}\label{sec:app}

\subsection{Positive energy conjectures}\label{ssec:pos_ene_theo}

Boundary conditions for general relativity that allow a well defined notion of total energy should have the property that there is a minimum energy solution consistent with those boundary conditions. The minimum energy solution is the gravitational ground state of the system. A series of positive energy theorems have been proven which show that standard boundary conditions indeed have such minimum energy solutions, which are often unique. For asymptotically AdS solutions with the usual boundary condition, $R\times S^{d-1}$, positive energy was first proven in four dimensions \cite{Gibbons:1982jg}, and later extended to higher dimensions \cite{Chrusciel:2003qr}.  Uniqueness of global AdS using the positive energy theorem was shown in four dimensions \cite{Boucher:1983cv} and extended to dimensions up to seven \cite{Qing}. 

\begin{figure}
    \centering
    \includegraphics[width=9cm]{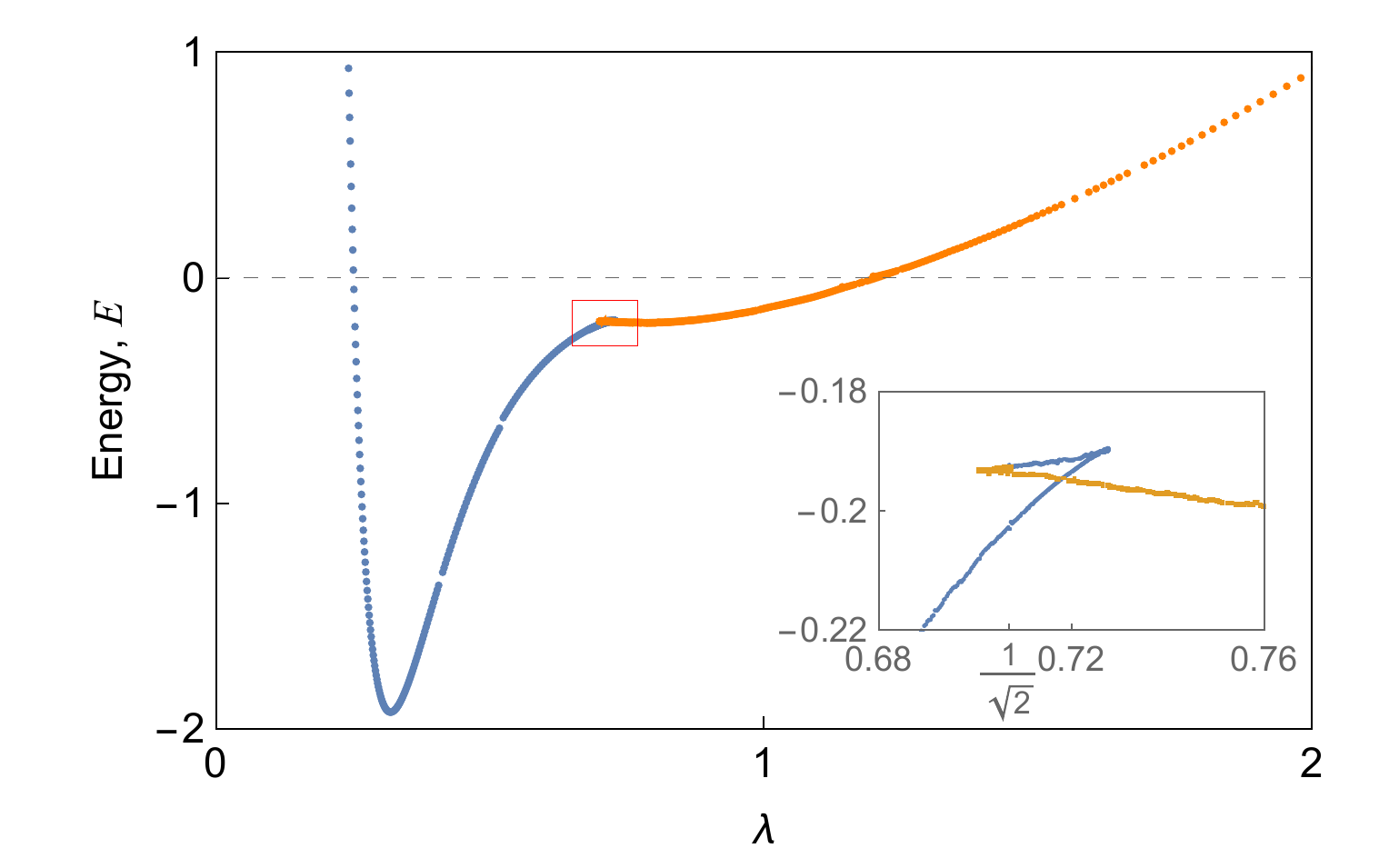}
    \caption{Energy for two families of ground state solutions in AdS${}_7$. The blue curve corresponds to solutions where $S^2$ shrinks to zero in the bulk and the orange curve corresponds to ones with $S^3$ shrinks to zero. The insert shows the energy of the first oscillation of solutions in Fig.~\ref{fig:AdS7_oscillation}. }
    \label{fig:AdS7_2SphereAction}
\end{figure}

We have computed the total energy of our solutions using the method described in App.~\ref{sec:RG}. We find that the lowest energy solution for each $\l$ is essentially given by the blue and orange curves in Fig.~\ref{fig:AdS7_oscillation} up to the point where they first have equal energy. All the oscillations at small $r_0$ result in solutions with higher energy. The lowest energy for each $\l$ is shown in  Fig.~\ref{fig:AdS7_2SphereAction}. Notice that the energy becomes large and positive for both large and and small $\l$, but is negative for $\l 
\sim \l_c$. The lowest energy $S^2$ contractible solutions meet the corresponding $S^3$ contractible solutions  at $\l = \l_t \approx 0.72$ which is slightly larger than $\l_c$.\footnote{{This matches  the value at which the Euclidean action of the two types of solutions agree when the conformal boundary is $S^2\times S^3$ without the time direction \cite{Aharony:2019vgs}.}}
The energy of some of the other solutions is shown in the insert. The solutions corresponding to smaller $r_0$ in Fig.~\ref{fig:AdS7_oscillation} have energies approaching $E = -5\pi^2/256$, in a narrow range around $\l_c$. This is because they approach the singular analytic solution as $r_0 \to 0$, which has this energy.

One expects the minimum energy solution to be static and highly symmetric \cite{Sudarsky:1992ty, Hertog:2005hm}. So our solution with lowest energy should be the ground state. This leads to new positive energy conjectures for asymptotically AdS gravity with boundary $R\times S^2 \times S^3 $:

\begin{conjecture}
Consider complete six dimensional initial data for the Einstein equation with $\Lambda < 0$, which has a conformal boundary $S^2 \times S^3$. Then its energy $E$ will satisfy $E\ge E_{min}$ where $E_{min}$ is the minimum energy shown in Fig.~\ref{fig:AdS7_2SphereAction}. Furthermore, the only solutions with $E=E_{min}$ are the ones we have constructed numerically.
\end{conjecture} 
 
This is really a one parameter family of conjectures, since it should hold for every ratio of the sphere radii, $\l$. Since we have found two solutions with the same minimum energy when $\l = \l_t$, if this conjecture is true, then one has an unusual situation where the ground state is not unique. 
 
Similar conjectures can be made in other dimensions. We construct the solutions with $S^2 \times S^2$ boundaries in App.~\ref{sec:S2S2}. 

The existence of a positive energy conjecture would have implications for holography, and we discuss some of them in Sec.~\ref{ssec:quan_PT} and Sec.~\ref{ssec:map}.

\subsection{Quantum phase transition}\label{ssec:quan_PT}

We now turn to consequences for a dual six dimensional CFT on $R\times S^2\times S^3$ using gravitational holography.  In holography, it is important that there is a positive energy theorem in the bulk, so the dual CFT has a stable vacuum state. The lowest energy solutions discussed above are  dual to the vacuum state of the boundary CFT at this $\l$. The value of the energy can be interpreted as a Casimir energy.  It is clear from Fig.~\ref{fig:AdS7_2SphereAction} that the Casimir energy depends on $\l$, and becomes large when one sphere is much bigger than the other.

A quantum phase transition is one that happens at zero temperature. To check whether our theory predicts a quantum phase transition, we only need to consider the lowest energy, zero-temperature solutions discussed above. (Since all our black holes have a finite temperature, they do not enter this discussion.) 

As can be seen from Fig.~\ref{fig:AdS7_2SphereAction}, the bulk dual of the vacuum state changes qualitatively at $\l=\l_t$ from one where the $S^2$ is contractible to one where
the $S^3$ is contractible. Since there is a discontinuity in the gradient as the two curves cross, this predicts a first-order quantum phase transition for the boundary CFT as $\l$ is varied. In other words, for a holographic CFT on $S^2\times S^3$, the vacuum depends on the relative size of the spheres and changes qualitatively at $\l=\l_t$.

\subsection{Canonical ensemble}
In a canonical ensemble, one fixes the temperature $T$ and looks for the state that minimizes the free energy $F\equiv-T\ln Z =E-TS$. Using a saddle point approximation, the partition function is given by $Z=e^{-I}$, where $I$ is the Euclidean action for the bulk saddle. Therefore, the free energy can be easily computed from the action via $F=-T\ln Z=I/\beta$, where $\beta$ is the periodicity of the Euclidean time, now equated to the inverse temperature. 

\begin{figure}[ht]
    \centering
    \begin{subfigure}{.48\linewidth}
		\includegraphics[width=8cm]{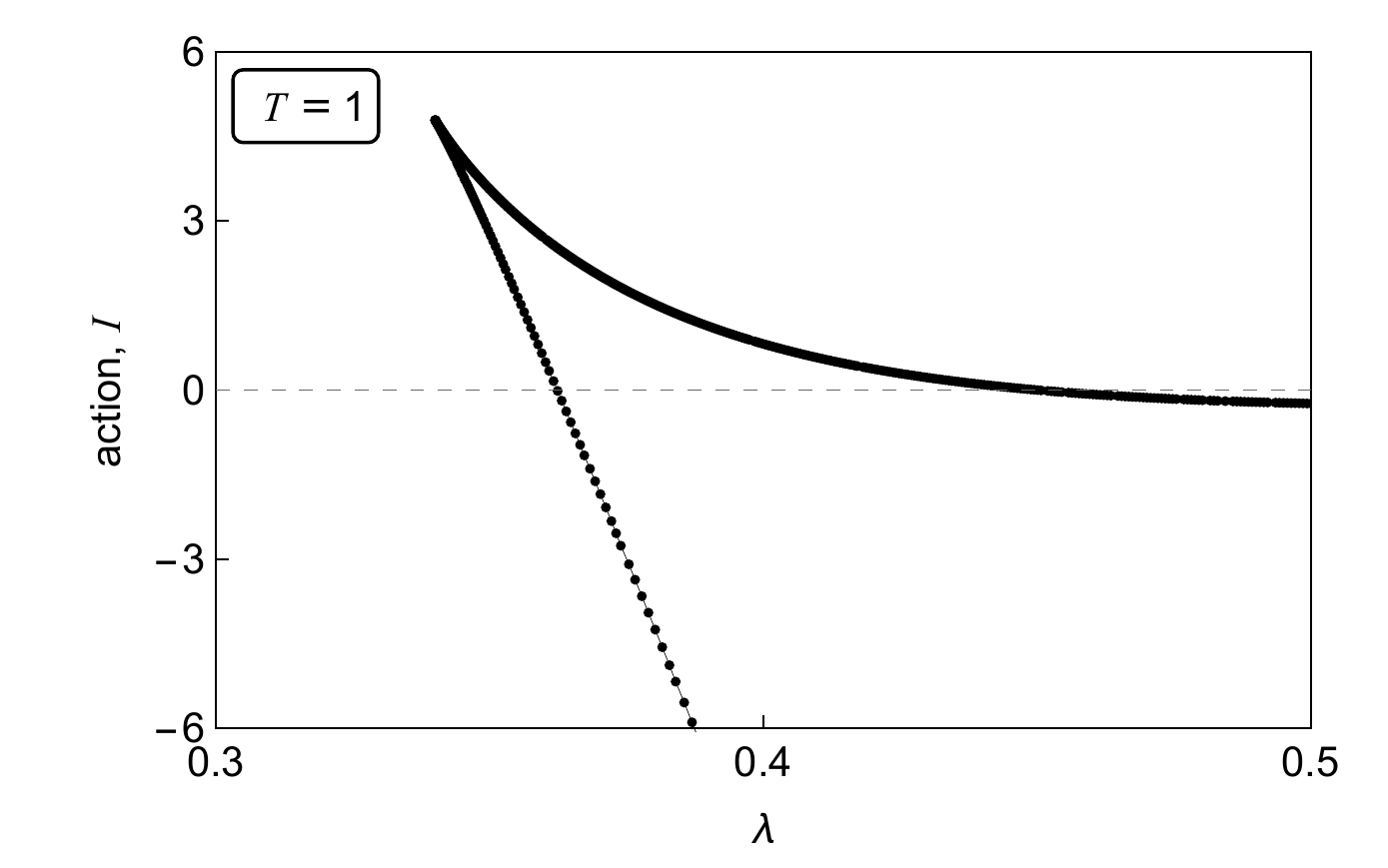} 
	\end{subfigure}
    \hskip1em
    \begin{subfigure}{.48\linewidth}
    	\includegraphics[width=8cm]{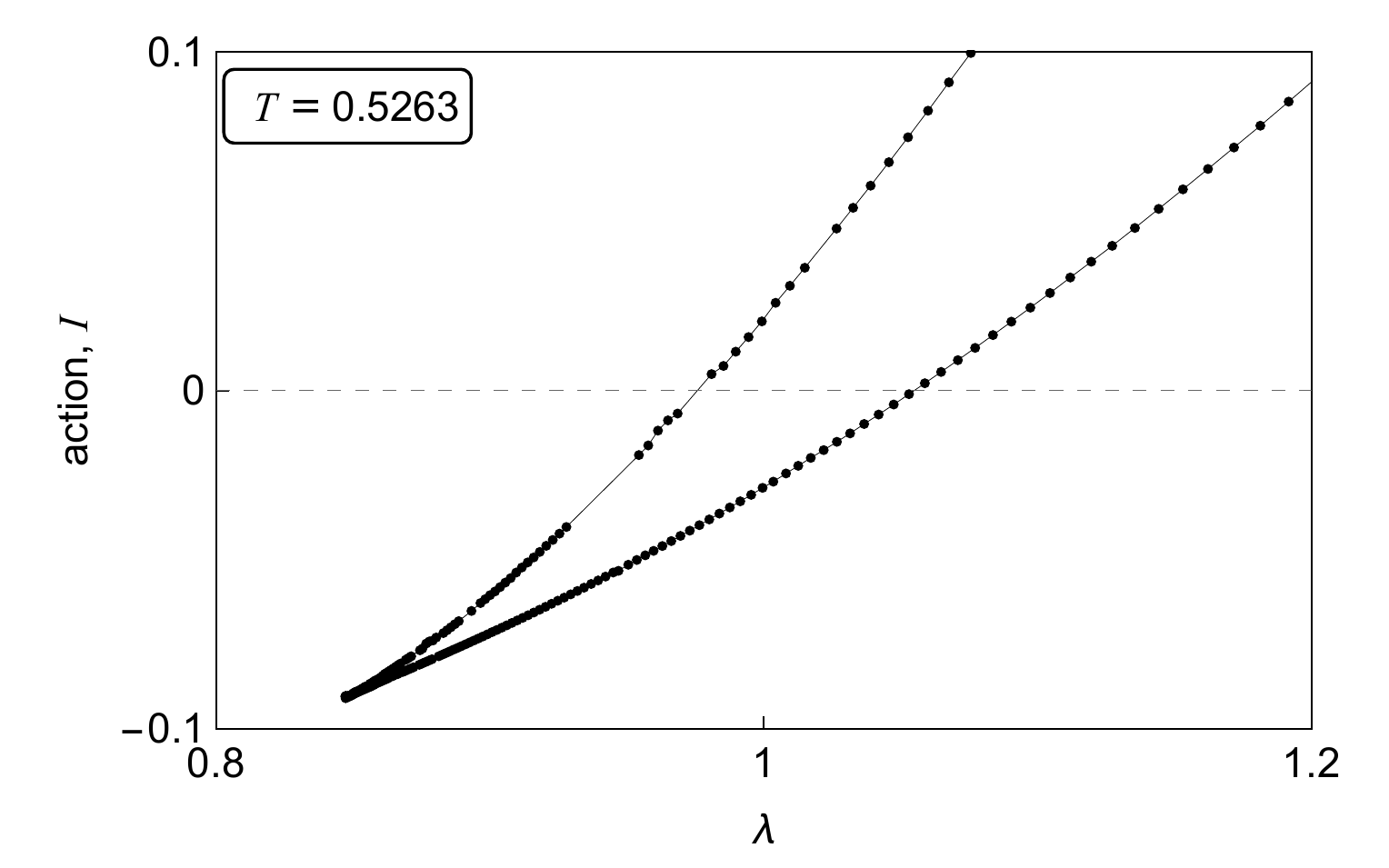}
    \end{subfigure}
    \caption{The Euclidean action of the black holes as a function of $\l$ at two different temperatures. At each temperature, there is a minimum value of $\l$ below which there can be no black holes. Similarly, for each $\l$, the black hole solution stops existing below a certain temperature, which can be seen from the fact that the tip of the curves moves to the right as temperature is lowered. There is also a global minimum temperature ($T\approx0.5$)  below which no black hole exists for any $\l$. These features are also visible in the phase diagram Fig.~\ref{fig:AdS7_Phase}.}
    \label{fig:AdS7_BhAction}
\end{figure}

We have computed the renormalized Euclidean action using the method described in App.~\ref{sec:RG}.
We show how the action depends on $\l$ in Fig.~\ref{fig:AdS7_BhAction} for two different temperatures. Notice that the behavior is very different in the two plots. On the left we show the behavior at high $T$ where the action for the large black hole becomes very negative. The upper curve shows the action for the small black hole. The other branches of small black holes that appear near $\l_c$ have actions that are close to this upper curve and never dominate the canonical ensemble.
On the right of Fig.~\ref{fig:AdS7_BhAction}, we consider a temperature close to the minimum allowed  black hole temperature. This is not possible for small $\l$ since the $S^2$ will pinch off before the $S^1$ when $\l$ is small. (In fact, a global minimum temperature for black holes exists since for low $T$, the boundary $S^1$ becomes so large that the $S^3$ pinches off before the $S^1$.) We see that for $T$ near its minimum value, the action now grows with $\l$ for both large and small black holes.

\begin{figure}[ht]
    \centering
    \includegraphics[width=9cm]{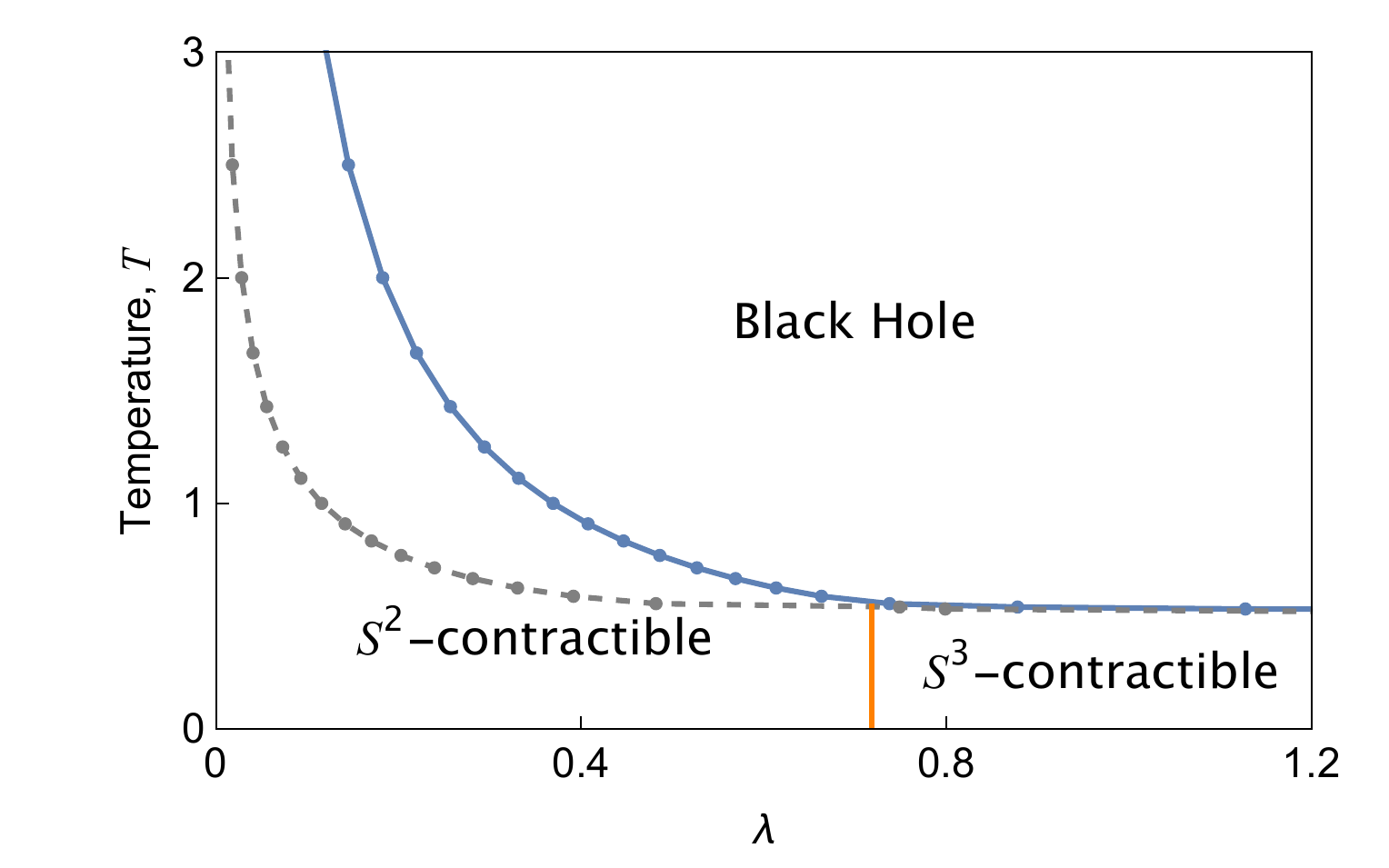}
    \caption{Phase diagram for the canonical ensemble  with conformal boundary $S^1\times S^2\times S^3$.  The blue line demarcates a transition between black holes and non-black holes, and the orange line separates two ground states. The gray dashed curve is the minimum temperature curve below which there are no black holes.}
    \label{fig:AdS7_Phase}
\end{figure}

By comparing the actions of the black hole solutions with a gas of gravitons on the non-black hole solutions, we can construct the phase diagram. This is shown in Fig.~\ref{fig:AdS7_Phase}. As we can see, the quantum phase transition extends vertically to some finite temperature. At each $\l$, there is a Hawking-Page phase transition, whose critical temperature depends on $\l$. According to the figure, the Hawking-Page phase transition seems to happen above the minimum temperature for a black hole solution to exist. The curve of minimum temperature asymptotes to some finite value around $1/2$, which is the global minimum temperature across all $\l$.

\subsection{Microcanonical ensemble}

\begin{figure}[ht]
    \centering
    \includegraphics[width=9cm]{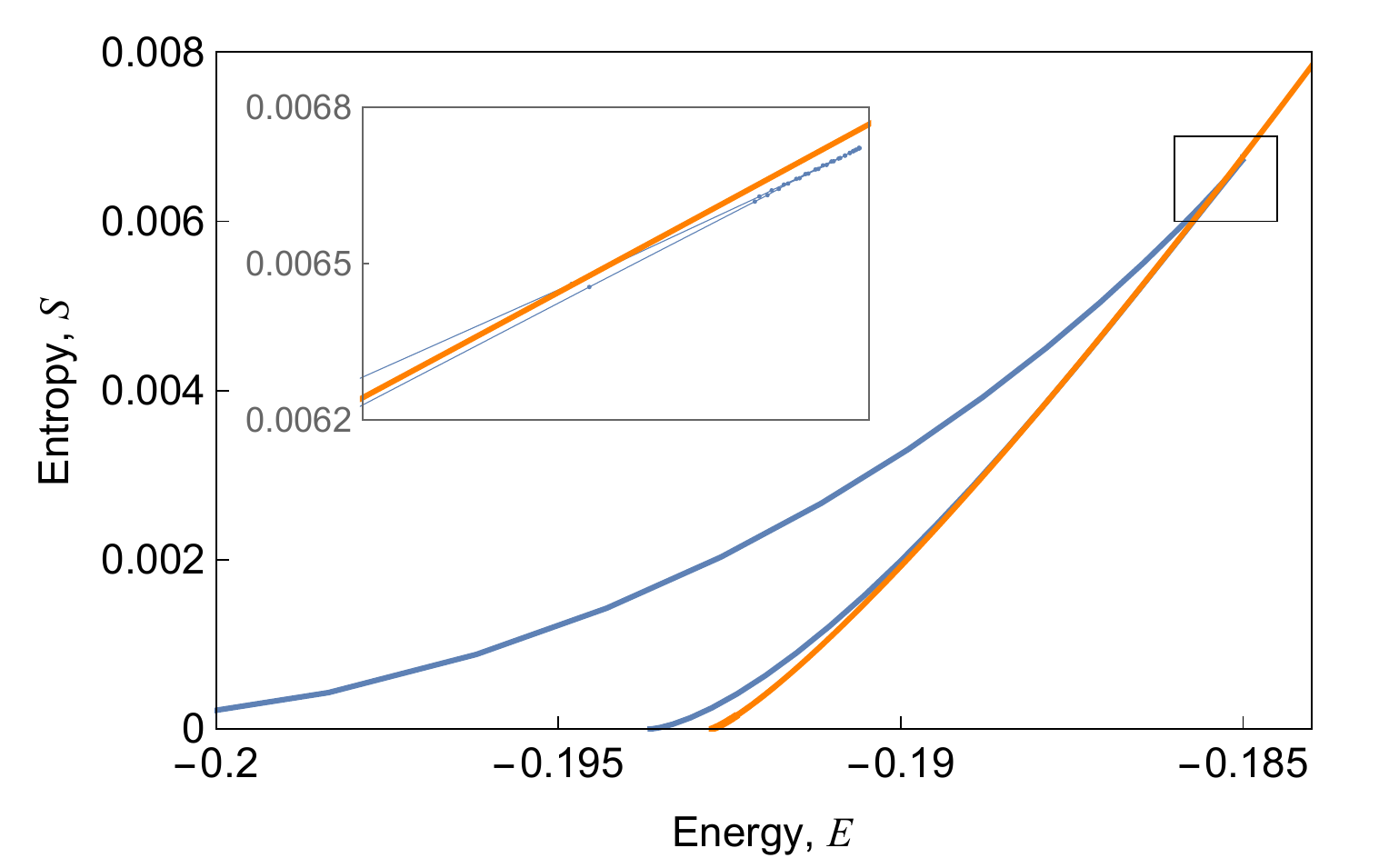}
    \caption{Entropy-energy plot for the analytic black hole solutions (shown in orange) and the first numerical branch of black holes (shown in blue) at the critical $\l=\sqrt{0.5}$. The insert shows that they cross at some finite $E$, so the analytic solutions stop being dominant when the energy is lowered below that value.  }
    \label{fig:micro}
\end{figure}

In a microcanonical ensemble, the energy is fixed, and the state with the largest entropy will be the most likely configuration. In holography, since entropy is given by $S=A/4$ (at leading order in $G$), where $A$ is the area of the horizon, for any given energy, a black hole will always be preferred over non-black hole states. In the usual story where the boundary is a $(d-1)$-sphere cross time, there is only one black hole at any given energy (above the ground state), so there is no competition (at leading order in $G$). In our case we have multiple candidates -- an important ingredient for a microcanonical phase transition.

In Fig.~\ref{fig:micro}, we plot the energy on the horizontal axis and entropy ($A/4$) on the vertical for the special $\l = \l_c$. At any fixed energy, we take the candidate with the largest entropy. Restricting to the set of solutions we are considering, the figures show that the analytical branch dominates at large energy. However, as energy is lowered, a numerical branch enters and intersects with it. This means that the analytic branch will stop being dominant. 

The other numerical branches have much smaller $r_0$ and require much more accuracy to plot, but we know some of their limits. First of all, in the zero-entropy limit, the solution reduces to a non-black hole solution. From Fig.~\ref{fig:AdS7_2SphereAction} and Fig.~\ref{fig:AdS7_oscillation}, the infinite oscillations go up in energy with smaller $r_0$, which translates to the fact that smaller black holes (or the smaller numerical branches) start at a larger value on the horizontal axis in Fig.~\ref{fig:micro}. The smaller branches also have smaller maximum area, so they will not reach as high as the first numerical branch along the vertical direction. Therefore, we expect these to lie within the envelope of the first numerical branch. Since we only take the one with maximum entropy, they will not compete with the first numerical branch.

In conclusion, when $\l = \l_c$ there is a change in dominance between the analytic branch and the first numerical branch in the microcanonical ensemble, and we may describe this as a `microcanonical  phase transition'. We expect a similar phase transition for $\l$ close to $\l_c$.

However, as the entropy approaches zero, one of the spheres on the horizon becomes much smaller than the other. In this regime, one expects a  Gregory-Laflamme instability, and a new branch of black holes with $S^5$ topology will have greater entropy with the same energy.
 These solutions will break the $SO(3)\times SO(4)$ symmetry and we have not studied them in detail. It is likely that they dominate the microcanonical ensemble only at low energy, so even if the above crossover between $S^2\times S^3$ black holes turns out to be subdominant to $S^5$ black holes, there will still be a crossover between the $S^5$ black holes and the analytic $S^2\times S^3$ black holes at higher energy.

\subsection{Confining gauge theories on de Sitter}

As another application of the Euclidean solutions, let us remark that these solutions can be analytically continued in more than one way to obtain different Lorentzian solutions. Instead of analytically continuing the thermal circle to Lorentzian time, we can analytically continue one of the  spheres to de Sitter space. The resulting bulk solution will now have a conformal boundary containing a de Sitter factor, and hence describe a quantum field theory in de Sitter space. This idea has been explored in e.g.~\cite{Aharony:2002cx,Balasubramanian:2002am,Birmingham:2002st,Ross:2004cb,Balasubramanian:2005bg,Astefanesei:2005eq,He:2007ji,Hutasoit:2009xy,Dibitetto:2020csn}. 

We now apply this analytic continuation to one of our boundary spheres, say the three-sphere. The CFT geometry $S^1\times S^2\times S^3$ then gets analytically continued to $S^1\times S^2\times \text{dS}_3$, where $S^1$ should now be considered as a Scherk-Schwarz compactification rather than a thermal one, with anti-periodic boundary conditions for the fermions. Instead of a CFT on $S^1\times S^2\times \text{dS}_3$, the Scherk-Schwarz compactification allows one to study confining gauge theories on $S^2\times \text{dS}_3$. The phase diagram we obtained for $S^1\times S^2\times S^3$ now implies that there are three phases for such theories. The bulk saddle with a contractible $S^3$ now analytically continues to a topological black hole. This describes a deconfined plasma phase. The bulk saddle with a contractible $S^1$ analytically continues to a generalization of the bubble of nothing with an extra factor of $S^2$.\footnote{As pointed out in \cite{Horowitz:2019dym}, although this is often called a ``bubble of nothing" in AdS, this is really a misnomer. In asymptotically flat spacetime, although a boundary at large radius is $S^1\times S^{d-1}$, there is a clear distinction between the circle and the sphere since the circle remains a finite size and the sphere becomes large. So when the circle pinches off, there is a ``hole" in the asymptotically flat space. However in AdS, both the circle and sphere become large asymptotically, and there is no reason to prefer one over the other.  So if the circle pinches off, one still has a complete asymptotically hyperbolic surface without any holes. The key difference is just that the spacetime now has a minimal $S^3$ rather than a minimal $S^1$.} Since the Scherk-Schwarz circle is contractible in the bulk, this describes a confining phase. The analogue of these two types of saddles were studied in \cite{Marolf:2010tg}. However, we also have a third type of phase where $S^2$ is contractible. This analytically continues to a solution with topology $S^1_{SS}\times R^3\times \text{dS}_{3}$. This is also a confined phase, although the SS circle does not shrink in the bulk. The infrared floors of these confining bulk solutions have different dimensions. One may interpret them as having different numbers of degrees of freedom in the IR. Similar phase structures were also studied in \cite{Blackman:2011in} for $S^2 \times \text{dS}_n$ and $T^2 \times \text{dS}_n$. Ours is more like their latter case due to the presence of three phases, but in our case we have unequal dimensions for the two possible IR floors, leading to two distinct confining phases.

Instead of $S^3\to\text{dS}{}_3$, we can also analytically continue $S^2\to\text{dS}{}_2$. The same argument then implies that there are different phases for confining gauge theories on $S^3\times \text{dS}_2$. Similarly, we can apply the same argument for the $S^2\times S^2$ example in App.~\ref{sec:S2S2}.  We also expect this to hold in higher dimensions (both the sphere dimension and de Sitter dimension) and lower ones, i.e., for confining gauge theories on $S^1\times S^2\to S^1\times \text{dS}_{2}$ \cite{Copsey:2006br}.

\subsection{Non-existence of a state-operator map}\label{ssec:map}
Positive energy theorems are important in general relativity since they identify the minimum energy solutions for given boundary conditions. Holographically, in addition to providing vacuum energies in the dual CFT, they have further implications. We now discuss a possible implication of this for the state-operator map for CFTs on $R\times S^2\times S^3$.

A state-operator map, as the name suggests, gives a map between states of the Hilbert space and  operators. It exists for two-dimensional CFTs on arbitrary Riemann surfaces, but for CFTs in higher dimensions, the map is not always defined. For CFTs on $R^{d}$ ($d>2$), the map exists between local operators and states on a sphere $S^{d-1}$. However, for other topology, the story is less clear. In \cite{Belin:2018jtf}, the possibility of a state-operator map for three-dimensional CFTs on $T^3$ was studied. In this case, one can consider whether a line operator can be mapped to a state on $T^2$. As they argue, to have a valid state-operator map, a necessary condition is to produce the vacuum with a compact Euclidean manifold. 

Using holography, it was argued that the positive energy conjecture for boundaries with topology $T^3$ (which has the AdS$_4$ soliton \cite{Horowitz:1998ha} as the unique ground state) implies no state-operator correspondence for states on $T^2$. To see this, consider the AdS$_4$ soliton, which is a static Lorentzian solution with a global timelike Killing vector field with topology $R\times R^2 \times S^1$. The Euclideanization of the time-direction gives a non-compact solution with the same topology. The Euclidean geometry that prepares the state at $t=0$ is obtained by cutting the geometry in half, i.e., replacing $R$ with $R_-$. However, the boundary of this half  soliton is $R_-\times T^2$ which is the standard way to construct the CFT ground state and is not compact.  If the conjecture is correct, no other geometry has the same energy and therefore no compact Euclidean geometry can prepare the vacuum state.

We now argue that, with the new conjectured positive energy theorems for boundaries with spatial topology of $S^2\times S^3$, a similar argument would imply no state-operator correspondence for states on $S^2\times S^3$. According to the conjecture, the bulk ground state is prepared by cutting the solutions in half, which has topology $R_-\times R^3\times S^3$ (for $\l<\l_t$) or $R_-\times S^2\times R^4$ (for $\l>\l_t$). The boundary of either manifold is simply $R_-\times S^2\times S^3$ which is non-compact and prepares the CFT vacuum state. Since the bulk lowest-energy solution is unique (for $\l\ne\l_t$), there is no compact six-dimensional Euclidean manifold that prepares the vacuum state, therefore disproving the existence of such a state-operator map. 

In fact, taking the conjecture to the more general situation of a product of an arbitrary number of spheres, it would seem that the state-operator map is forbidden for a large class of topology in higher dimensional CFT. It would be interesting to further explore the relations between positive energy theorems and the existence of a state-operator map on spaces with more general topology.

\section{Discussion}\label{sec:disc}
We have studied Euclidean solutions in general relativity with a negative cosmological constant that have a conformal boundary $S^1\times S^2\times S^3$ with symmetry $U(1)\times SO(3)\times SO(4)$. Both black holes and non-black holes have been found, and the dependence of the number of solutions on the ratio $\l$ between the radii of the two spheres at infinity has been discussed. 

At each $\l$,  the non-black hole solution with lowest energy is considered to be the ground state. We then conjectured that this is the unique ground state among all bulk solutions with given boundary metric, except at $\l_t \approx 0.72$ where there are two solutions with the same  energy, in which case we conjecture that they are the only two ground states. A better understanding of this exceptional case may shed light on the uniqueness of ground states in general relativity. Moreover, it could be interesting to perform some numerical tests of these positive energy conjectures similar to those done in \cite{Copsey:2006br}. These non-black hole solutions have an interesting dependence on $\l$, in that the number of them becomes arbitrarily large as $\l$ goes to a critical ratio $\l_c = \sqrt {0.5}$. This generalizes the phenomenon found in \cite{Aharony:2019vgs} where the boundary metric is $S^2\times S^3$ without a time direction. Curiously, this only happens when the total dimension of the two spheres is smaller than nine. It would be interesting to investigate more generally the criteria for when the oscillations in Fig.~\ref{fig:AdS7_oscillation} can happen and for what topology.\footnote{A good starting point might be the spaces $T^{p,q}$ which are $S^1$ bundles over $S^2\times S^2$ and admit an Einstein metric. We thank Igor Klebanov for this suggestion.}

We next studied black hole solutions where we found that the number of one-parameter families of solutions depends on $\l$ and the number goes to infinity as $\l \to \l_c$. This demonstrates an extreme form of black hole non-uniqueness in higher dimensional general relativity even with no matter fields. There are many potential directions for its generalization: adding angular momentum and/or charge will make the phase structure more non-trivial, and allow one to investigate whether the extremal limits of the infinity of black holes exhibit unusual behavior.  One could also consider replacing the spheres by two compact negatively curved spaces. There is again a simple analytic solution when the product is an Einstein space. Are there infinitely many others?

We have not studied the stability of the black hole or non-black hole solutions, and this should be investigated. For the non-black hole solutions, at every extremum of $\l$ in Fig.~\ref{fig:AdS7_oscillation} a small change in $r_0$ does not change $\l$. This means there is a static zero mode.\footnote{We thank Toby Wiseman for pointing this out.} Usually a zero mode arises when a stable mode transitions to an unstable mode (or vice versa). Thus starting at large $r_0$ where there is only one solution for the given $\l$ and we expect it to be stable, the solutions between the first and second extrema are likely to be unstable. At the second extrema, we either add a second unstable mode or remove the unstable mode we had. So it is not  clear if the solutions between the second and third extrema are stable, but the solutions between the third and fourth extrema are probably unstable, etc. For the black hole solutions, we have already mentioned that they are unstable when one sphere on the horizon is much smaller than the other. However it is possible that there are additional instabilities.

The holographic implications of these bulk solutions were then investigated. These include a three-phase diagram for the CFT in a thermal ensemble and for confining gauge theories on de Sitter, a phase transition in the microcanonical ensemble, and  no state-operator correspondence. Most of the infinite solutions never become dominant in either the canonical or microcanonical ensemble. With the usual boundary topology $R\times S^{d-1}$, going to the microcanonical ensemble places small black holes that do not dominate in the canonical ensemble in the spotlight \cite{Marolf:2018ldl}. However, in our case, there is just not enough spotlight for infinitely many saddles, even with the microcanonical ensemble. It is an interesting open question whether the infinity of solutions for a product of spheres plays a role in \textit{any} (holographic) gravitational path integral. If the answer if yes, then we can say that the CFT senses their presence in the bulk via some appropriate bulk computation. Said differently, one should try to understand whether these infinitely many non-dominating saddles mean anything in the CFT, other than provide exponentially small corrections.

Another feature worth mentioning is that the one-parameter family of Hawking-Page transitions all seem to happen above the minimum temperature for which black hole solutions exist (only checked numerically up to some large $\l$). Therefore, the free energy is desirably continuous at the transition which is then necessarily (at least) first order. This suggests that there may be some protection mechanism of the gravitational path integral that always make the thermal saddle take over before the temperature is lowered to the point where black holes are forbidden. It would be interesting to check whether this is true with more general boundary topology and/or extra matter fields. More generally, one may wonder whether this protection mechanism is ubiquitous in gravitational path integrals: saddles with different topology always switch dominance continuously in free energy. 

Finally, we comment on another potential direction. The boundary geometries have been chosen to be maximally symmetric for simplicity.  However, investigating how bulk quantities (in particular the energy  \cite{Cheamsawat:2019gho, Fischetti:2016vfq}) respond to perturbations in the shape of the boundary spheres can lead to important observations, so it might be worth exploring such perturbations in our case also. In particular, a large class of less symmetric Einstein metrics on $S^2\times S^3$ are now known (see e.g.~\cite{Cvetic:2005ft}), and could be the starting point of such investigations.

\section*{Acknowledgements} 
It is a pleasure to thank Don Marolf and Toby Wiseman for discussions. This work was supported in part by NSF Grant
PHY-2107939.

\appendix
\section{Argument for an infinity of non-black holes} \label{sec:pert}
{An infinity of solutions was found in \cite{Aharony:2019vgs} when the conformal boundary is a product of spheres without the extra time dimension. We show in this appendix that the argument generalizes to our case with the extra dimension, where the Euclidean solutions they found become static Lorentzian solutions, which we refer to as non-black holes.} We consider $S^m\times S^n$ where $m\le n$.  The singular non-black hole solution can be obtained by taking $r_0\to0$ in our family of analytic solutions \eqref{eq:AdSpq} followed by coordinate rescaling, giving
\begin{equation}
    ds^2 = - \mathcal{A}(r) dt^2 +\frac{dr^2}{\mathcal{A}(r)} + r^2 \left [\frac{m-1}{n-1}\ d\Omega_m^2   + d\Omega_n^2\right],
    \end{equation}
where 
\begin{equation}
    \mathcal{A}(r) = r^2 +\frac{n-1}{m+n-1}.
\end{equation}

Making a coordinate transformation
\begin{equation}
    r= \sinh z \sqrt{\frac{n-1}{m+n-1}},
\end{equation}
we move to the gauge similar to that used in \cite{Aharony:2019vgs}:
\begin{equation}
    ds^2=-g^2_s(z)dt^2+dz^2+f^2_s(z)d\Omega_m^2+h^2_s(z)d\Omega_n^2,
\end{equation}
where
\begin{equation}\begin{aligned}
    g_s(z)&=\sqrt{\frac{n-1}{m+n-1}}\cosh z,\\
    f_s(z)&=\sqrt{\frac{m-1}{m+n-1}}\sinh z,\\
    h_s(z)&=\sqrt{\frac{n-1}{m+n-1}}\sinh z.
\end{aligned}\end{equation}

Smooth solutions where one of the spheres pinches off at  $r_0 \ll 1$ will look very different near $r_0$, but we expect that for $z\gg r_0$, it will be just a small perturbation of this singular metric. So we start by studying these perturbations.
We consider the same perturbation as in \cite{Aharony:2019vgs}, except now we have an extra perturbation in time:
\begin{equation}
    \delta f(z)=\frac{1}{m} f_{s}(z) a(z), \quad \delta h(z)=-\frac{1}{n} h_{s}(z) a(z), \quad \delta g(z)=g_s(z)b(z),
\end{equation}
With this ansatz, the $S^m$ and $S^n$ components of the linearized Einstein equation give: 
\be 
\begin{aligned}\label{eq:EtiltSnSm}
   -\frac{(m-1) \sinh (2 z) b'(z)}{4 (b(z)-1) (m+n-1)}+\frac{m-1}{2 m (b(z)-1) (m+n-1)}\times
   \mathcal{F} = 0 \\
  -\frac{(n-1) \sinh (2 z) b'(z)}{4 (b(z)-1) (m+n-1)}+\frac{n-1}{2 n (b(z)-1) (m+n-1)}\times
  \mathcal{F} = 0.
\end{aligned}
\ee
where
\be 
\begin{aligned}
         \mathcal{F}  =   &a'(z) \left[\sinh ^2(z) b'(z)+(b(z)-1) ((m+n+1) \sinh (2 z)-2 \tanh (z))\right]\\
            +&2 (b(z)-1) \sinh ^2(z) a''(z)+a(z) \left(\sinh (2 z) b'(z)+4 (b(z)-1) (m+n-1)\right).
    \end{aligned}
\ee
Eliminating $a(z)$, the equations give $b^\prime(z)=0$. This implies the perturbation in $g_{tt}$ is just a multiple of itself, which can be redefined away by rescaling $t$. We therefore set $b(z)=0$. Eliminating $b(z)$ (or setting it to zero), equations \eqref{eq:EtiltSnSm} reduce to
\begin{equation}\label{eq:Etilt}
    \sinh ^2(z) a''(z)+ \left[(m+n+1) \sinh (z) \cosh (z)-\tanh (z)\right]a'(z)+2 (m+n-1) a(z)=0,
\end{equation}
which is solved by
\begin{equation}\label{eq:apm}
\begin{aligned}
    a_\pm(z)=\,_2F_1\left[
       \frac{1}{2}\alpha_\pm,
       \frac{1}{2}\alpha_\pm;
       \alpha_\pm+\frac{1}{2}\left(m+n+1\right);
       \tanh ^2(z)\right]\tanh^{\alpha_\pm}(z),
\end{aligned}\end{equation}
where $\alpha_\pm\equiv-\frac{1}{2}\left(m+n-1\pm\sqrt{(m+n-1)(m+n-9)}\right)$. We next repeat the argument in \cite{Aharony:2019vgs} to show the existence of infinitely many oscillations. When  $m+n<9$, $a_\pm$ are complex and  a general real solution can be written as $a(z)=(a_0\cdot a_+(z)+a_0^*\cdot a_{-}(z))$ where $a_0$ is a complex constant. The asymptotic ratio of the sphere radii, $\l$, can be worked out directly from the metric and is given by   $\l = \l_c + \delta \l$ where
\begin{equation}
\frac{\delta \l}{\l_{c}}=\left(\frac{1}{m}+\frac{1}{n}\right)\left(a_{0} \cdot a_+(\infty)+a_0^*\cdot a_-(\infty)\right).
\end{equation}

Without loss of generality, consider now the smooth nonlinear solution where $S^m$ shrinks to a point before $S^n$ so the topology of spacetime is $R^{m+2}\times S^n$ (or $S^1\times R^{m+1}\times S^n$ in Euclidean signature) and define $r_0$ to be the radius of the minimal $S^n$. For $r_0$ much smaller than the AdS length, the solution around $z\sim r_0$ is determined by $r_0$, the only length scale in this region. However, for $z\gg r_0$, the solution resembles a small perturbation of the singular solution. Define $a_0(r_0)$ to be the value of $a_0$ such that the metric due to the linear perturbation $a(z)$ behaves similar to the solution with small $r_0$ for $z\gg r_0$. From \eqref{eq:apm}, for $r_0\ll z\ll 1$, 
\begin{equation}
    a_\pm(z)\sim z^{\alpha_\pm} \implies a_0(r_0)\sim r_0^{-\alpha_\pm},
\end{equation}
where the dependence on $r_0$ follows from the fact that $a(z)$ is dimensionless and hence can only depend on the dimensionless ratio $z/r_0$. Therefore,
\begin{equation}
    \delta \lambda(r_0) \sim r_0 ^\frac{m+n-1}{2} \cos\left(\sqrt{(m+n-1)(9-m-n)}\log r_0+\phi\right),
\end{equation}
for some phase $\phi$. This shows that there are an infinite number of oscillations in Fig.~\ref{fig:AdS7_oscillation}.

\section{Numerical procedure for solving the Einstein equation}\label{sec:num_ODE}
In this appendix we provide an algorithm for  finding the solutions numerically. Following \cite{Copsey:2006br}, we begin by rewriting the components of the Einstein equation in a more user-friendly form. In AdS${}_7$, the Einstein equation is given by
\begin{equation}
    G_{ab}=R_{ab}-\frac{1}{2}g_{ab}R=-\L g_{ab}={15}\,g_{ab}.
\end{equation}

\subsection{\texorpdfstring{$S^2$}{S2} contractible}\label{ssec:S2n}
For $S^2$-contractible solutions, we take the ansatz \eqref{AdS7_S2metric}. From the $G_{tt}$ component, we have
\begin{equation}\begin{aligned}\label{AdS7_Gtt}
   &6 \mathcal{C} \left(-10 r^2+r \mathcal{B}
   \mathcal{A}'+\mathcal{A} \left(r \mathcal{B}'+2
   \mathcal{B}\right)-2\right)+2 r \mathcal{C}'
   \left(r \mathcal{B} \mathcal{A}'+\mathcal{A}
   \left(r \mathcal{B}'+6
   \mathcal{B}\right)\right)\\
   &=r^2 \left(\frac{\mathcal{A} \mathcal{B}
   \left(\mathcal{C}'^2-4 \mathcal{C}
   \mathcal{C}''\right)}{\mathcal{C}}+\frac{4}{\lambda ^2}\right),
\end{aligned}\end{equation}
and the component on the three-sphere (which we will denote $G_{S^3}$) gives us
\begin{equation}\begin{aligned}\label{AdS7_GS3}
   &\mathcal{C} \left(-60 r^2+2r^2\mathcal{B}\mathcal{A}''+\mathcal{A}' \left(r^2\mathcal{B}'+8r \mathcal{B}\right)+4\mathcal{A} \left(r\mathcal{B}'+\mathcal{B}\right)-4\right)\\
   &=r^2 \left(\frac{\mathcal{A} \mathcal{B}\left(\mathcal{C}'^2-4 \mathcal{C}\mathcal{C}''\right)}{\mathcal{C}}+\frac{4}{\lambda ^2}\right)-2 r \mathcal{C}' \left(2 r \mathcal{B}\mathcal{A}'+\mathcal{A} \left(r \mathcal{B}'+4\mathcal{B}\right)\right).
\end{aligned}\end{equation}
Subtracting \eqref{AdS7_Gtt} from \eqref{AdS7_GS3} allows us to solve for $\mathcal{C}^\prime$:
\begin{equation}\label{AdS7_Cprime}
    \mathcal{C}'=-\frac{\mathcal{C} \left(r^2
   \mathcal{A}' \mathcal{B}'+2 r \mathcal{B}
   \left(\mathcal{A}'+r \mathcal{A}''\right)-2
   \mathcal{A} \left(r \mathcal{B}'+4
   \mathcal{B}\right)+8\right)}{2 r \mathcal{B}
   \left(r \mathcal{A}'-2 \mathcal{A}\right)}.
\end{equation}
Next, the component on the two-sphere gives us
\begin{equation}\begin{aligned}\label{AdS7_GS2}
   &\mathcal{C}^2 \left( 2 r^2\mathcal{B}
   \mathcal{A}''+ r^2\mathcal{A}'
   \mathcal{B}'+12r \mathcal{B}\mathcal{A}'+ 6r\mathcal{A} \mathcal{B}'+12\mathcal{A}
   \mathcal{B}-12\right)-r^2\mathcal{A}
   \mathcal{B} \mathcal{C}'^2\\
   &=r\mathcal{C} \left(60r
   \mathcal{C}-2r
   \mathcal{A} \mathcal{B} \mathcal{C}''-\mathcal{C}' \left(2 r
   \mathcal{B} \mathcal{A}'+ r\mathcal{A}
   \mathcal{B}'+6\mathcal{A} \mathcal{B}\right)\right),
\end{aligned}\end{equation}
and the $G_{rr}$ component is given by
\begin{equation}\label{AdS7_Grr}
   \frac{r\mathcal{B}\mathcal{C}'\left(2\mathcal{C} \left(6 \mathcal{A}+r\mathcal{A}'\right)+r\mathcal{A}\mathcal{C}'\right)}{\mathcal{C}}=6\mathcal{C}\left(10r^2-\mathcal{B}\left(2\mathcal{A}+r\mathcal{A}'\right)+2\right)+\frac{4r^2}{\lambda^2}.
\end{equation}
Subtracting \eqref{AdS7_GS2} twice and \eqref{AdS7_Grr} once from \eqref{AdS7_Gtt}, together with the substitution of $\mathcal{C}^\prime$ from \eqref{AdS7_Cprime}, we obtain an expression for $\mathcal{B}$ in terms of $\mathcal{A}$ and its derivatives only:
\begin{equation}\label{AdS7_beta}
    \mathcal{B}=\frac{2 \left(\mathcal{A}'+3 r^2
   \mathcal{A}'-6 r \mathcal{A}
   \right)}{-\mathcal{A} \mathcal{A}'-r\mathcal{A}\mathcal{A}''+r\mathcal{A}'^2}.
\end{equation}
Next, substitute \eqref{AdS7_beta} and its derivatives with respect to $r$ into the Einstein equation. From \eqref{AdS7_Gtt}, we will get an expression for $\mathcal{A}^{\prime\prime\prime}$ in terms of $\mathcal{A}$ and $\mathcal{C}$ up to their second derivatives. From \eqref{AdS7_GS2}, we substitute $\mathcal{A}^{\prime\prime\prime}$ we just derived and obtain an expression for $\mathcal{A}^{\prime\prime}(r,\mathcal{A},\mathcal{A}^\prime,\mathcal{C},\mathcal{C}^\prime,\mathcal{C}^{\prime\prime})$. From \eqref{AdS7_Grr}, with the substitution of $\mathcal{A}^{\prime\prime}$ and  $\mathcal{A}^{\prime\prime\prime}$, we will obtain $\mathcal{A}^\prime(r,\mathcal{A},\mathcal{C},\mathcal{C}^\prime,\mathcal{C}^{\prime\prime})$.
Taking the derivative of $\mathcal{A}^\prime$ and equating it to $\mathcal{A}^{\prime\prime}(r,\mathcal{A},\mathcal{A}^\prime,\mathcal{C},\mathcal{C}^\prime,\mathcal{C}^{\prime\prime})$ with substitution of $\mathcal{A}^\prime$ returns a third order non-linear ODE for $\mathcal{C}$ {(for $r>r_0$)}:
\begin{equation}\begin{aligned}\label{AdS7_eqC}
    &\,\left(3\lambda^2\mathcal{C} \left(5 r^2+1\right)+r^2\right)\times\\
    &\left[\begin{aligned}
        &r^3 \mathcal{C}'^4 \left(3\lambda^2r^2\mathcal{C}'+\lambda^2\mathcal{C}'-3 r\right)+36 \lambda^2r\mathcal{C}^4\left(3\mathcal{C}'+r\mathcal{C}''-r^2\mathcal{C}^{(3)}\right)\\
        +\,&r^2\mathcal{C} \mathcal{C}'^2\left( 15\lambda^2r^2\mathcal{C}'^2+11\lambda^2\mathcal{C}'^2+3 r\mathcal{C}''\left( 3\lambda^2r^2\mathcal{C}'+\lambda^2\mathcal{C}'+r\right)-15 r\mathcal{C}'\right)\\
        +\,&2r\mathcal{C}^2\left(\begin{aligned}
            &\,r^3 \mathcal{C}''^2+3 \lambda^2\mathcal{C}'^3\left(1-12 r^2\right)\\
            +\,&\lambda ^2 \mathcal{C}'^2\left( 3r^4\mathcal{C}^{(3)}+r^2\mathcal{C}^{(3)}-3 r^3\mathcal{C}''-4r\mathcal{C}''\right)\\
            -\,&r^2\mathcal{C}'\left( 3\lambda^2\mathcal{C}''^2\left(3 r^2+1\right)-11\mathcal{C}''+r\mathcal{C}^{(3)}\right)
        \end{aligned}\right)\\
        +\,&6 \mathcal{C}^3\left(\begin{aligned}
            &\,r^2 \left(\mathcal{C}''-2 \lambda^2\mathcal{C}''^2 \left(2r^2+1\right)-r\mathcal{C}^{(3)}\right)-3 \lambda^2\mathcal{C}'^2\left(3 r^2+1\right)\\
            +\,&r \mathcal{C}' \left(\lambda^2\mathcal{C}^{(3)}\left(r^3+r\right)+\lambda^2\mathcal{C}''\left(13 r^2-3\right)+3\right)
        \end{aligned}\right)
    \end{aligned}\right]=0.
\end{aligned}\end{equation}

The boundary conditions for solving \eqref{AdS7_eqC} are as follows.  Since we have a third order equation, we have to specify $\MC$ and its first two derivatives at one point $r_0$. We set $\MC(r_0)=0$ so $r_0$ is the point where the $S^2$ pinches off. {We are only interested in solutions that are not singular at $r_0$, so if \eqref{AdS7_eqC} holds for $r > r_0$, it must also hold at $r_0$. Evaluating it at $r_0$ gives a condition on $C'(r_0)$, and evaluating its derivative here gives a condition on $C''(r_0)$. These boundary conditions then determine $\MC(r)$ for all $r$.} {Requiring $\MC(r)\to r^2$ at large $r$ further picks out a unique $\lambda$.}\footnote{{It may be useful to define $\widetilde\MC=\lambda^2\MC$ in \eqref{AdS7_eqC}, in which case $\l$ is read off from the asymptotic behaviour of $\widetilde\MC= \l^2 r^2+\cdots$.}} In this way, we obtain a unique solution for each choice of $r_0$.

Once we solved for $\MC$, it is numerically convenient to solve for $\MA$ using a second order coupled differential equation of $\MC$ and $\MA$, obtained as follows. First, obtain $\MB(r,\MA,\MA^\prime,\MC,\MC^\prime)$ through eliminating $\MA^{\prime\prime}$ and $\MC^{\prime\prime}$ using \eqref{AdS7_Gtt}, \eqref{AdS7_GS2} and \eqref{AdS7_Grr}. Then, substitute the expression of $\MB(r,\MA,\MA^\prime,\MC,\MC^\prime)$ and its derivative with respect to $r$ into \eqref{AdS7_Gtt} to obtain $\MA^{\prime\prime}(r,\MA,\MA^\prime,\MC,\MC^\prime,\MC^{\prime\prime})$. Finally, substituting the expression of $\MB$, $\MB^\prime$ and $\MA^{\prime\prime}$ we derived from previous steps into \eqref{AdS7_GS2} gives
\begin{equation}\begin{aligned}\label{AdS7_eqcoupled}
    &\,2 \mathcal{C}\mathcal{A}'\left(3\lambda^2r^2\mathcal{C}'+\lambda^2\mathcal{C}'-6\lambda^2r\mathcal{C}-r\right)\\
    +\,&\frac{\mathcal{A}}{r^2 \mathcal{C}'+3 r\mathcal{C}}\times\left(\begin{aligned}
        &\,3\lambda^2r^4\mathcal{C}'^3+\lambda^2r^2\mathcal{C}'^3-3r^3\mathcal{C}'^2-72\lambda^2r \mathcal{C}^3\\
        +\,&2r\mathcal{C}\left(3\lambda^2\mathcal{C}'^2+r^2\mathcal{C}''-5r\mathcal{C}'\right)\\
        +\,&6\lambda^2\mathcal{C}^2\left(\mathcal{C}''\left(5r^3+r\right)-\mathcal{C}'\left(r^2-3\right)\right)-12r\mathcal{C}^2
    \end{aligned}\right)=0.
\end{aligned}\end{equation}
{Note that this is a linear, first order ODE for $\MA(r)$.}
Requiring $\MA(r)\to r^2$ for large $r$ and substituting the numerically obtained $\MC(r)$ into \eqref{AdS7_eqcoupled} gives the numerical answer for $\MA(r)$.

\subsection{\texorpdfstring{$S^3$}{S3} contractible}\label{ssec:S3n}
For $S^3$-contractible solutions, we take the ansatz \eqref{AdS7_S3metric}. Obtaining the equation for $\mathcal{C}$ in the $S^3$-contractible case is very similar to what we did for $S^2$-contractible solutions. Subtracting the $G_{tt}$ component from $G_{S_2}$ allows us to solve for $\mathcal{C}^\prime$:
\begin{equation}\label{AdS7_S3c_C1p}
    \mathcal{C}'=\frac{\mathcal{C} \left(\lambda ^2 \left(2 \mathcal{A} \left(r \mathcal{B}'+2 \mathcal{B}\right)-r^2\left(2 \mathcal{B} \mathcal{A}''+\mathcal{A}' \mathcal{B}'\right)\right)-4\right)}{3 \lambda ^2 r \mathcal{B}\left(r \mathcal{A}'-2 \mathcal{A}\right)}.
\end{equation}
Next, subtracting three copies of $G_{S_3}$ and one copy of $G_{rr}$ from two copies of $G_{tt}$, together with the substitution of $\mathcal{C}^\prime$ given in \eqref{AdS7_S3c_C1p} allow us to solve for $\mathcal{B}$:
\begin{equation}\label{AdS7_S3c_beta}
    \mathcal{B}=-\frac{\mathcal{A}'+6\lambda^2r^2\mathcal{A}'-12\lambda^2r\mathcal{A}}{\lambda^2\left(\mathcal{A}\mathcal{A}'+r\mathcal{A}\mathcal{A}''-r\mathcal{A}'^2\right)}.
\end{equation}
Then, substitute \eqref{AdS7_S3c_beta} back into the Einstein equation, and eliminate $\mathcal{A}^{\prime\prime\prime}$, $\mathcal{A}^{\prime\prime}$ and $\mathcal{A}^\prime$ with steps similar to App.~\ref{ssec:S2n}  returns again a third order ODE for {$\mathcal{C}(r>r_0)$}:
\begin{equation}\label{AdS7_S3c_eqC}\begin{aligned}
    &\,\left(\mathcal{C}+3\lambda^2r^2+15\lambda^2r^2\mathcal{C}\right)\times\\
    &\left[\begin{aligned}
        &\,3r^3\mathcal{C}'^4\left(\mathcal{C}'+6\lambda^2r^2\mathcal{C}'-8\lambda^2r\right)-48\lambda^2r\mathcal{C}^4\left(r^2\mathcal{C}^{(3)}-2\mathcal{C}'\right)\\
        +\,&r^2\mathcal{C}\mathcal{C}'^2\left(9\mathcal{C}'^2+3r\mathcal{C}''\left(6\lambda^2r^2\mathcal{C}'+\mathcal{C}'+8\lambda^2r\right)-18\lambda^2r^2\mathcal{C}'^2-40\lambda^2r\mathcal{C}'\right)\\
        +\,&4\mathcal{C}^3\times\left(\begin{aligned}
           &\,2\mathcal{C}'^2\left(6\lambda^2r^2-1\right)-\mathcal{C}''^2\left(3\lambda^2r^4+2r^2\right)-4\lambda^2r^3\mathcal{C}^{(3)}\\
           +\,&r \mathcal{C}' \left(8 \lambda^2+\mathcal{C}^{(3)} \left(r-3 \lambda ^2r^3\right)+\mathcal{C}'' \left(39 \lambda^2r^2-2\right)\right)
        \end{aligned}\right)\\
        +\,&r\mathcal{C}^2\times\left(\begin{aligned}
            \,&12\lambda^2r^3\mathcal{C}''^2-4\mathcal{C}'^3\left(36\lambda^2r^2+1\right)\\
            +\,&r\mathcal{C}'^2\left(32\lambda^2+3\mathcal{C}^{(3)}\left(6\lambda^2r^3+r\right)+\mathcal{C}''\left(6\lambda^2r^2-11\right)\right)\\
            -\,&r^2\mathcal{C}'\left(-68\lambda^2\mathcal{C}''+9\mathcal{C}''^2\left(6\lambda^2r^2+1\right)+12\lambda^2r\mathcal{C}^{(3)}\right)
        \end{aligned}\right)\end{aligned}\right]=0.
\end{aligned}\end{equation}
Again, {as in Sec.~\ref{ssec:S2n}}, setting $\mathcal{C}\left(r_0\right)=0$, evaluating \eqref{AdS7_S3c_eqC} {and its derivative} at $r=r_0$ and requiring $\mathcal{C}\to r^2$ for large $r$ {picks out a unique $\l$} for each $r_0$. Once we solved for $\mathcal{C}$, we follow steps similar to App.~\ref{ssec:S2n} to solve for $\mathcal{A}$. Through eliminating $\mathcal{A}^{\prime\prime}$, $\mathcal{B}$ and $\mathcal{B}^{\prime}$, we obtain a first order equation for  $\mathcal{A}(r)$:
\begin{equation}\begin{aligned}\label{AdS7_S3c_eqcoupled}
    &\,r\mathcal{C}\mathcal{A}'\mathcal{C}'\left(4\mathcal{C}+3r\mathcal{C}'\right)\left(\mathcal{C}'\left(6\lambda^2r^2+1\right)-4\lambda^2r(3\mathcal{C}+1)\right)\\
    +\,&\mathcal{A}\mathcal{C}'\times\left(\begin{aligned}
        &\,4r\mathcal{C}\left(3\lambda^2r^2\mathcal{C}''+\mathcal{C}'^2\left(2-6\lambda^2r^2\right)-9\lambda^2r\mathcal{C}'\right)\\
        +\,&4 \mathcal{C}^2\left(\mathcal{C}''\left(15\lambda^2r^3+r\right)+\mathcal{C}'\left(2-15\lambda^2r^2\right)-4\lambda^2r\right)\\
        +\,&3r\left(r\mathcal{C}'^2\left(6\lambda^2r^2\mathcal{C}'+\mathcal{C}'-8\lambda^2r\right)-16\lambda^2\mathcal{C}^3\right)
        \end{aligned}\right)=0,
\end{aligned}\end{equation}
which has a unique solution with $\mathcal{A}(r)\to r^2$.

\subsection{Black hole}\label{ssec:num_BH}
The ansatz we used to solve for the black hole solutions \eqref{AdS7_BHmetric} is identical to the $S^2$-contractible one, except that we now use Euclidean signature. This will not change the behavior of the Einstein equation, so we can use all the equations in App.~\ref{ssec:S2n}.  

Substituting \eqref{AdS7_Cprime}, \eqref{AdS7_beta} and its derivative into $G_{rr}$ allows us to solve for $\mathcal{C}$ in terms of $\mathcal{A}$:
\begin{equation}\begin{aligned}\label{AdS7_gamma}
    \lambda^2\mathcal{C}=&-8r^2\left(\mathcal{A}'\left(3 r^2+1\right)-6r\mathcal{A}\right) \left(\mathcal{A}\left(\mathcal{A}'+r\mathcal{A}''\right)-r\mathcal{A}'^2\right)^3\div\\
    &\left\{\begin{aligned}
        -&8r^3\left(3r^2+1\right)^2\mathcal{A}'^6 \left(3\mathcal{A}'+r\mathcal{A}''\right)-1692r^2\mathcal{A}^5\mathcal{A}'^2\\
        +&36r^2\mathcal{A}^5\left(r^4\mathcal{A}^{(3)2}-15r^2\mathcal{A}''^2+2r^2\mathcal{A}^{(3)}\left(7\mathcal{A}'+9r\mathcal{A}''\right)-66r\mathcal{A}'\mathcal{A}''\right)\\
        +&4r^2\mathcal{A}\mathcal{A}'^4 \left[\begin{aligned}
            \,&6r^2\mathcal{A}''^2\left(3 r^2+1\right)^2+3\mathcal{A}'^2\left(99r^4+32r^2+1\right)\\
            +&\mathcal{A}'\left(3r\mathcal{A}''\left(51r^4+32r^2+5\right)-\mathcal{A}^{(3)}\left(3r^3+r\right)^2\right)\end{aligned}\right]\\
        +&8r\mathcal{A}^2\mathcal{A}'^2\left[\begin{aligned}
            \,&3\mathcal{A}'^3\left(-57 r^4+7r^2+2\right)-3r^3\mathcal{A}''^3\left(3r^2+1\right)^2\\
            -&r\mathcal{A}'^2\left(r\mathcal{A}^{(3)}\left(3r^2+1\right)+\mathcal{A}''\left(324r^4+81r^2-1\right)\right)\\
            +&r^2\mathcal{A}'\mathcal{A}''\left(3 r^2+1\right) \left(r\mathcal{A}^{(3)}\left(3r^2+1\right)-2\mathcal{A}''\left(9r^2+2\right)\right)
            \end{aligned}\right]\\
        +&12r\mathcal{A}^4 \left[\begin{aligned}
            \,&18r^2\mathcal{A}'\mathcal{A}''^2\left(2-5r^2\right)+6\mathcal{A}'^3\left(52r^2+7\right)\\
            +&3r^3\mathcal{A}''^2\left(r\mathcal{A}^{(3)}\left(3r^2+1\right)+\mathcal{A}''\left(7r^2+5\right)\right)\\
            +&r\mathcal{A}'^2\left(\mathcal{A}''\left(141r^2+67\right)-r\mathcal{A}^{(3)}\left(81r^2+13\right)\right)\\
            +&r^2\mathcal{A}'\left(-r^2\mathcal{A}^{(3)2}\left(3r^2+1\right)-2r\mathcal{A}''\mathcal{A}^{(3)}\left(24r^2+7\right)\right)
            \end{aligned}\right]\\
        +&\mathcal{A}^3\left[\begin{aligned}
            \,&2r^3\mathcal{A}'^2\mathcal{A}''\mathcal{A}^{(3)}\left(-9r^4+12r^2+5\right)\\
            +&9r^4\mathcal{A}''^4\left(3r^2+1\right)^2-12\mathcal{A}'^4\left(138r^4+76r^2+3\right)\\
            -&6r^3\mathcal{A}'\mathcal{A}''^2\left(3 r^2+1\right) \left(r\mathcal{A}^{(3)}\left(3r^2+1\right)+\left(r^2+1\right)\mathcal{A}''\right)\\
            +&r^2\mathcal{A}'^2\left(r^2\mathcal{A}^{(3)2}\left(3r^2+1\right)^2+\mathcal{A}''^2\left(1161r^4+30r^2-35\right)\right)\\
            +&12r\mathcal{A}'^3\left(r\mathcal{A}^{(3)}\left(42r^4+17r^2+1\right)+\mathcal{A}''\left(228r^4-41 r^2-5\right)\right)
            \end{aligned}\right]
    \end{aligned}\right\}.
\end{aligned}\end{equation}
Now, with substitution of \eqref{AdS7_beta} and its derivative, we can equate \eqref{AdS7_Cprime} and the derivative of \eqref{AdS7_gamma} to obtain a fourth order ODE for $\mathcal{A}(r)$. We can solve this fourth order ODE directly to obtain $\MA(r)$ and use \eqref{AdS7_gamma} to obtain $\MC(r)$. The boundary conditions needed are $\MA(r_0)$, $\MA'(r_0)$, $\MA''(r_0)$ and $\MA^{(3)}(r_0)$, so together with $r_0$ there are five independent parameters. However, they must satisfy four constraints. First, $\MA(r_0) = 0$ and
requiring $\MA(r)\to r^2$ removes two degrees of freedom. Also, generically $\MA^{(4)}(r_0)$ is divergent, and requiring regularity there removes another degree of freedom by putting a constraint on the combination of boundary conditions, which can be found by evaluating the ODE at $r_0$, giving
\begin{equation}\label{AdS7_a3pr0}
    \mathcal{A}^{(3)}\left(r_0\right)=\frac{5 \left(1-3 r_0^2\right) \mathcal{A}''\left(r_0\right)}{3 \left(3r_0^3+r_0\right)}+\frac{\left(27 r_0^4+20 r_0^2+5\right) \mathcal{A}'\left(r_0\right)}{\left(3r_0^3+r_0\right){}^2}+\frac{4 \mathcal{A}''\left(r_0\right){}^2}{3 \mathcal{A}'\left(r_0\right)}.
\end{equation}
At this point, two degrees of freedom remain, which can be identified as temperature and $r_0$. However, since $\MC(r)$ is determined from $\MA(r)$ according to \eqref{AdS7_gamma}, generic temperature and $r_0$ do not lead to the correct asymptotic behavior for $\MC(r)$, and requiring $\MC(r)\to r^2$ removes a further degree of freedom. Therefore, for fixed $\l$, there is only one degree of freedom. 

Although we have presented a complete algorithm for obtaining all the solutions, in practice, solving a fourth order ODE leads to large numerical error. We now describe a method with better accuracy by dealing with $\MC(r)$ first. According to \eqref{AdS7_eqC}, one can obtain $\MC(r)$ by solving a third order ODE. The function $\MA(r)$ is then obtained by solving the first order ODE \eqref{AdS7_eqcoupled}, with $\MC(r)$ known numerically. The constraints on the boundary conditions can be obtained as follows. 

Evaluate \eqref{AdS7_eqcoupled} at the location of horizon $r=r_0$ to obtain a constraint on $\mathcal{C}(r_0)$:
\begin{equation}\label{AdS7_BHCr0}
    r_0+6\lambda^2r_0\mathcal{C}(r_0)-\lambda^2\left(3r_0^2+1\right)\mathcal{C}'(r_0)=0.
\end{equation}
Then, evaluating \eqref{AdS7_gamma} at $r_0$ gives us:
\begin{equation}\label{AdS7_a2pr0}
    \mathcal{A}''\left(r_0\right)=-\frac{\left(9 r_0^2\mathcal{C}\left(r_0\right)+3\mathcal{C}\left(r_0\right)+r_0^2\right) \mathcal{A}'\left(r_0\right)}{\left(3 r_0^3+r_0\right)\mathcal{C}\left(r_0\right)}.
\end{equation}
Finally, take the derivative of \eqref{AdS7_Cprime} and evaluate it at $r=r_0$. Using \eqref{AdS7_a3pr0}, \eqref{AdS7_a2pr0}, \eqref{AdS7_beta} and its derivative gives us the constraint:
\begin{equation}\label{eq:AdS7_BHCpp}
    \mathcal{C}''(r_0)=\frac{\left(6r_0\mathcal{C}(r_0)+r_0\right)^2}{2\left(3r_0^2+1\right)^2\mathcal{C}(r_0)}.
\end{equation}
In terms of the degrees of freedom, for any given $r_0$, we see that $\MC'(r_0)$ and $\MC''(r_0)$ are fixed in terms of $\MC(r_0)$ according to \eqref{AdS7_BHCr0} and \eqref{eq:AdS7_BHCpp} respectively. So the only degree of freedom (for any given $r_0$) is $\MC(r_0)$, but this is removed by requiring $\MC(r)\to r^2$. For $\MA$, there is no degree of freedom remaining because $\MA(r)\to r^2$ uniquely fixed the solution due to it being first order. Therefore, the only thing one can vary is $r_0$, and of course $\l$ (which changes the ODEs). The temperature comes out as an output since it is fixed by $\MA'(r_0)$.

\section{Numerical holographic renormalization}\label{sec:RG}

For asymptotically AdS spacetimes, the renormalized Euclidean action is given by\footnote{We use the  convention $R^{a}{}_{b c d}=\left(\Gamma^{a}{}_{b d,c}+\Gamma^{e}{}_{b d} \Gamma^a{}_{e c}\right)-(c \leftrightarrow d)$ which differs from some of the references.} \cite{Henningson:1998gx,Emparan:1999pm,deHaro:2000vlm}:
\begin{equation}\label{total_action}
    I=\lim_{r_c\to\infty}\left[I_\text{bulk}(r_c)+I_\text{bdry}(r_c)+I_\text{CT}(r_c)\right],
\end{equation}
where
\begin{equation}
    I_\text{bulk}(r_c)=
        -\frac{1}{2\kappa}\int_{\mathcal{M}_c}\sqrt{g}\left(R-2\Lambda\right)
\end{equation}
is the Einstein-Hilbert action with a cosmological constant evaluated up to a radial cutoff, $r_c$,
\begin{equation}
    I_\text{bdry}(r_c)= -\frac{1}{\kappa}\int_{\partial\mathcal{M}_c}\sqrt{h}\,K 
\end{equation}
is the Gibbons-Hawking-York boundary action required for a good variational principle, evaluated \textit{on} the cut-off surface, and
\begin{equation}
    \begin{aligned}
        \,I_\text{CT}(r_c)
        =\,\frac{1}{\kappa}\int_{\partial\mathcal{M}_c}\sqrt{h}&\left[(d-1)+\frac{1}{2(d-2)}\mathcal{R}\right.\\
        &\left.+\frac{1}{2(d-4)(d-2)^2}\left(\mathcal{R}_{ab}\mathcal{R}^{ab}-\frac{d}{4(d-1)}\mathcal{R}^2\right)+\cdots\right]
    \end{aligned}
\end{equation}
is the counter term required to cancel the divergences near the AdS boundary, also evaluated on the cut-off surface. Here, $\mathcal{R}$ and $\mathcal{R}_{ab}$ are the Ricci scalar and Ricci tensor for the induced metric $h_{ab}$ on the cut-off surface, and $R$ is the Ricci scalar for the bulk metric $g_{ab}$. The dots in the counter terms represent terms not needed in dimensions less than 8. We have also used $\kappa=8\pi G_N$ and $K=\nabla_{a}n^a$ with $n^a$ being the outward pointing normal vector of the boundary. 

The number of terms needed in practice depends on the dimension. For AdS${}_6$, the terms shown are sufficient, but for AdS${}_7$, an additional term is required to cancel potential logarithmic divergences:
\begin{equation}
    \begin{aligned}
        I_\text{CT}^\text{log}(r_c)=-\frac{1}{\kappa}\int_{\partial\mathcal{M}_c}\sqrt{h}&\left[\frac{1}{64}\left(\frac{1}{2}\mathcal{R}\mathcal{R}_{ab}\mathcal{R}^{ab}-\frac{3}{50}\mathcal{R}^3-\mathcal{R}_{ac}\mathcal{R}_{bd}\mathcal{R}^{abcd}\right.\right.\\
        &\left.\left.+\frac{1}{5}\mathcal{R}^{ab}D_aD_b\mathcal{R}-\frac{1}{2}\mathcal{R}^{ab}\square\mathcal{R}_{ab}+\frac{1}{20}\mathcal{R}\square \mathcal{R}\right)\log(r_c)\right].
    \end{aligned}
\end{equation}
For this expression to be exact, we just need $g_{rr}\sim 1/r^2$ at leading order for large $r$. This is related to the Fefferman-Graham radial coordinate $z$ where $g_{zz}=1/z^2$ by $r=1/z$ plus subleading corrections, which all go to zero as $r_c\to\infty$.

To obtain the renormalized energy for the ground states (non-black hole solutions), we just need to replace $\mathcal{M}_c$ by $\Sigma_c$ which is any constant-$t$ hypersurface with $r_c>0$ removed. To see this, notice that energy is related to the Euclidean action by $E=I/\beta$. Since static non-black hole solutions do not have dependence on $\beta$, division by $\beta$ is equivalent to excluding the time integral in the expressions. 

One technical hurdle in performing this renormalization numerically is that infinities do not cancel out due to large numerical errors. Following \cite{Aharony:2019vgs}, we tame this as follows. First, we rewrite the surface terms (both boundary and counter terms) as bulk integrals:
\begin{equation}
    I_\text{bdry}+I_\text{CT} = \int^{r_c}\partial_r(I_\text{bdry}+I_\text{CT})dr.
\end{equation}
This turns the full action into an integral on $\mathcal{M}_c$. Since the total action \eqref{total_action} is finite, the integrand of $\int dr$ cannot diverge, either. The trick then is to obtain a finite integrand before performing the $r$-integration. At large $r$, the integrand decays to zero. One can therefore cut off the tail of the integrand at some $r$ for large enough $r$. For better accuracy, one can use the fact that the integrand decays with a specific power (worked out by substituting the asymptotic series expansion of the metric into \eqref{total_action}). The coefficient of this leading power can be found from matching to the numerical solution. Therefore, above some large $r$, one can replace the integrand by an expression that can be integrated analytically. Either way, the problem of numerical errors building up at large $r$ is circumvented.

\section{A lower-dimensional example: \texorpdfstring{$S^2\times S^2$}{S2 times S2}}\label{sec:S2S2}

\begin{figure}[ht]
    \centering
    \includegraphics[width=9cm]{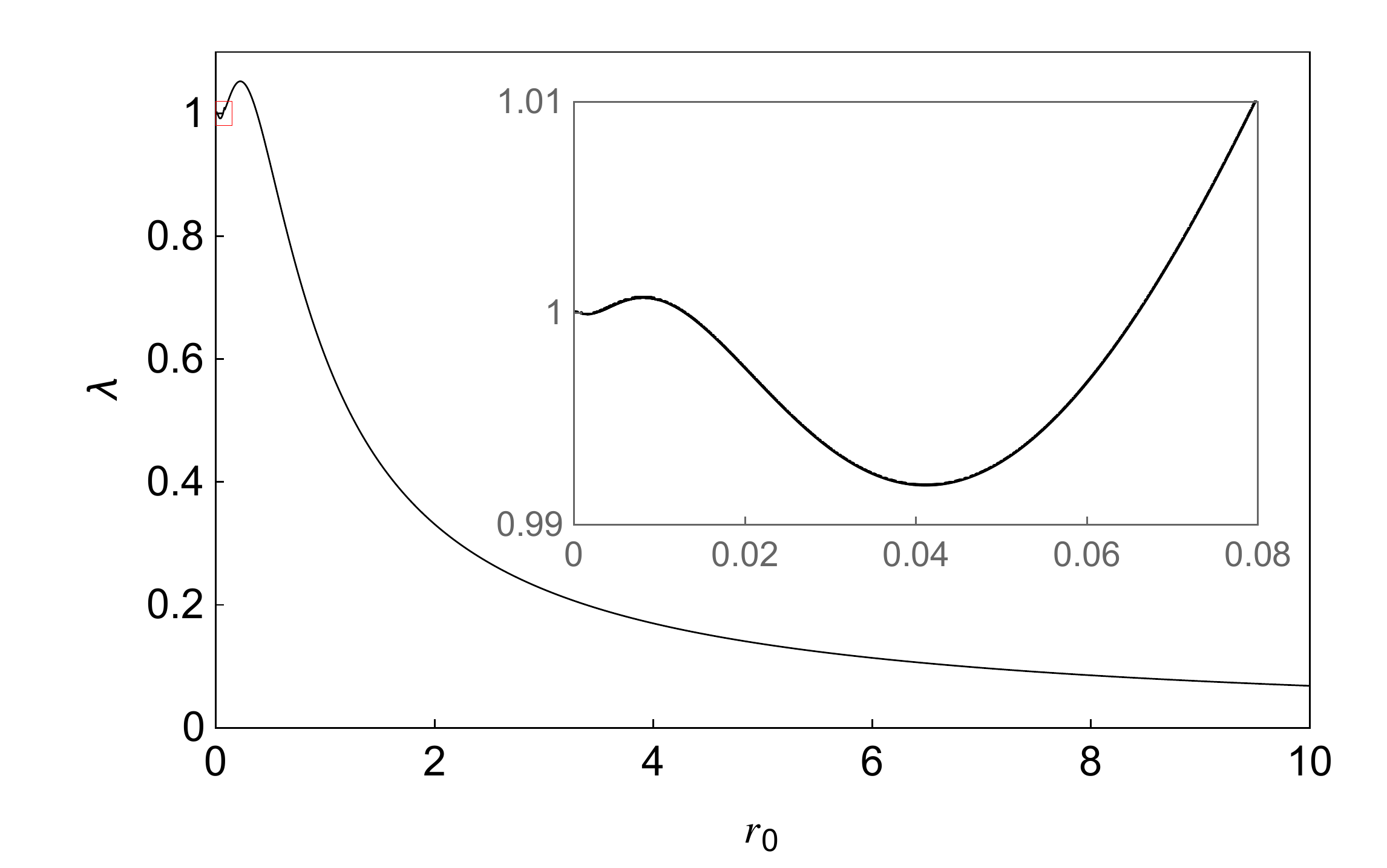}
    \caption{Non-black hole solutions in AdS$_6$ with the first sphere contractible.  There are an infinite number of oscillations for small $r_0$ and $\l \approx 1$. The insert shows the first two.}
    \label{fig:AdS6_oscillation}
\end{figure}
It is not uncommon for physical phenomena to occur in even but not odd number of dimensions or vice versa. The details of the interesting results discussed in the main sections of the paper certainly depend on the dimension, but as a first step in generalizing them, we investigate whether the main features occur for odd-dimensional boundary CFTs by looking at an example in one lower dimension, namely AdS$_6$ with conformal boundary $R\times S^2\times S^2$:
\begin{equation}
\label{eq:g_bdy_S2S2}
    ds^2|_{\partial \mathcal{M}}= - dt^2 + \l^2 d\Omega_2 + d\tilde\Omega_2.
\end{equation}

Conveniently, given that both spheres are of the same dimension, we can take the first sphere to pinch off without loss of generality (for non-black hole solutions).  We again find an oscillatory behaviour for these non-black hole solutions, as shown in Fig.~\ref{fig:AdS6_oscillation}. The oscillations are about the critical ratio for $S^2\times S^2$, which by symmetry is $\l=\l_c =1$.
\begin{figure}[ht]
    \centering
    \begin{subfigure}{.48\linewidth}
    	\includegraphics[width=8cm]{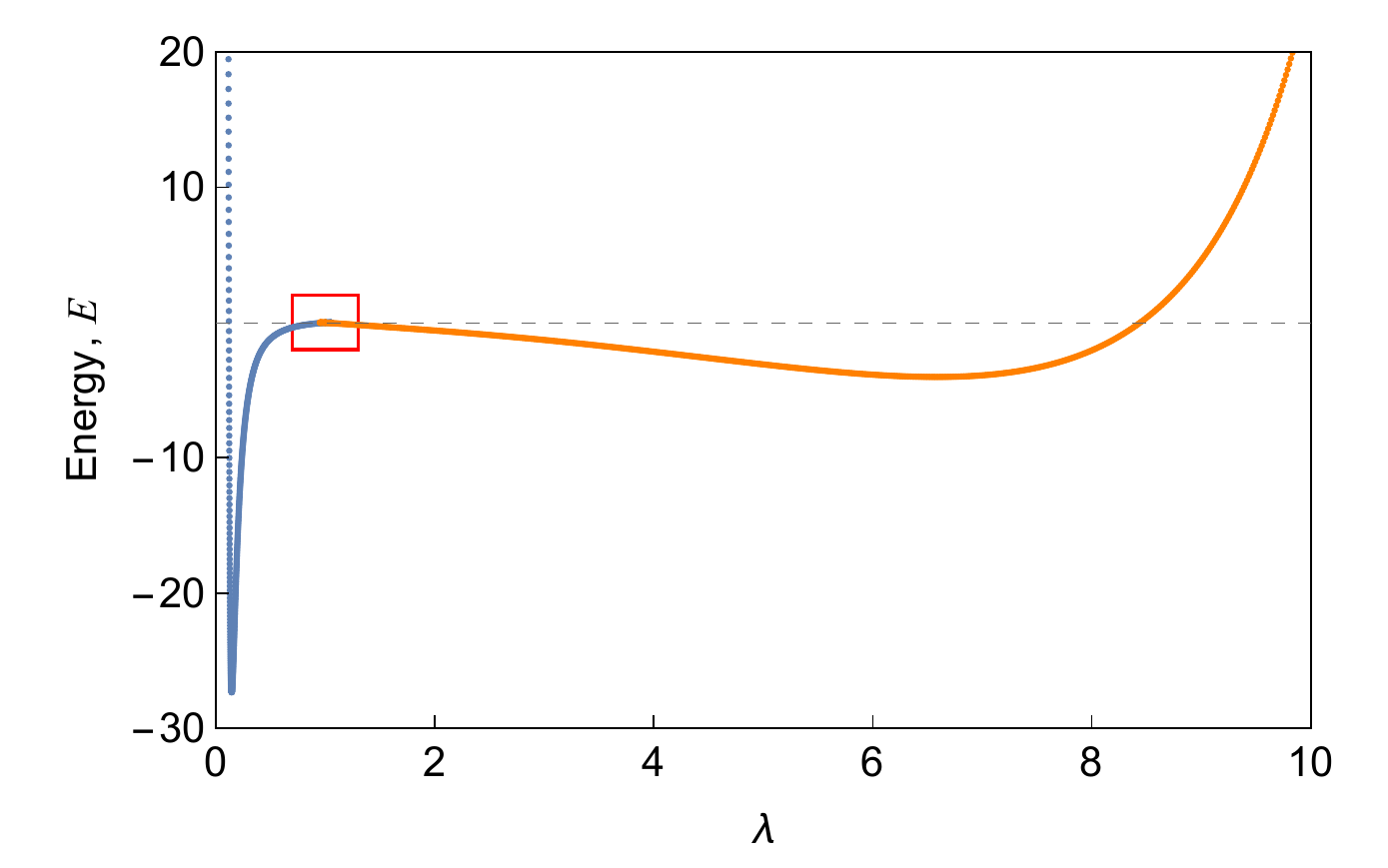}
    \end{subfigure}
    \hskip1em
    \begin{subfigure}{.48\linewidth}
    	\includegraphics[width=8cm]{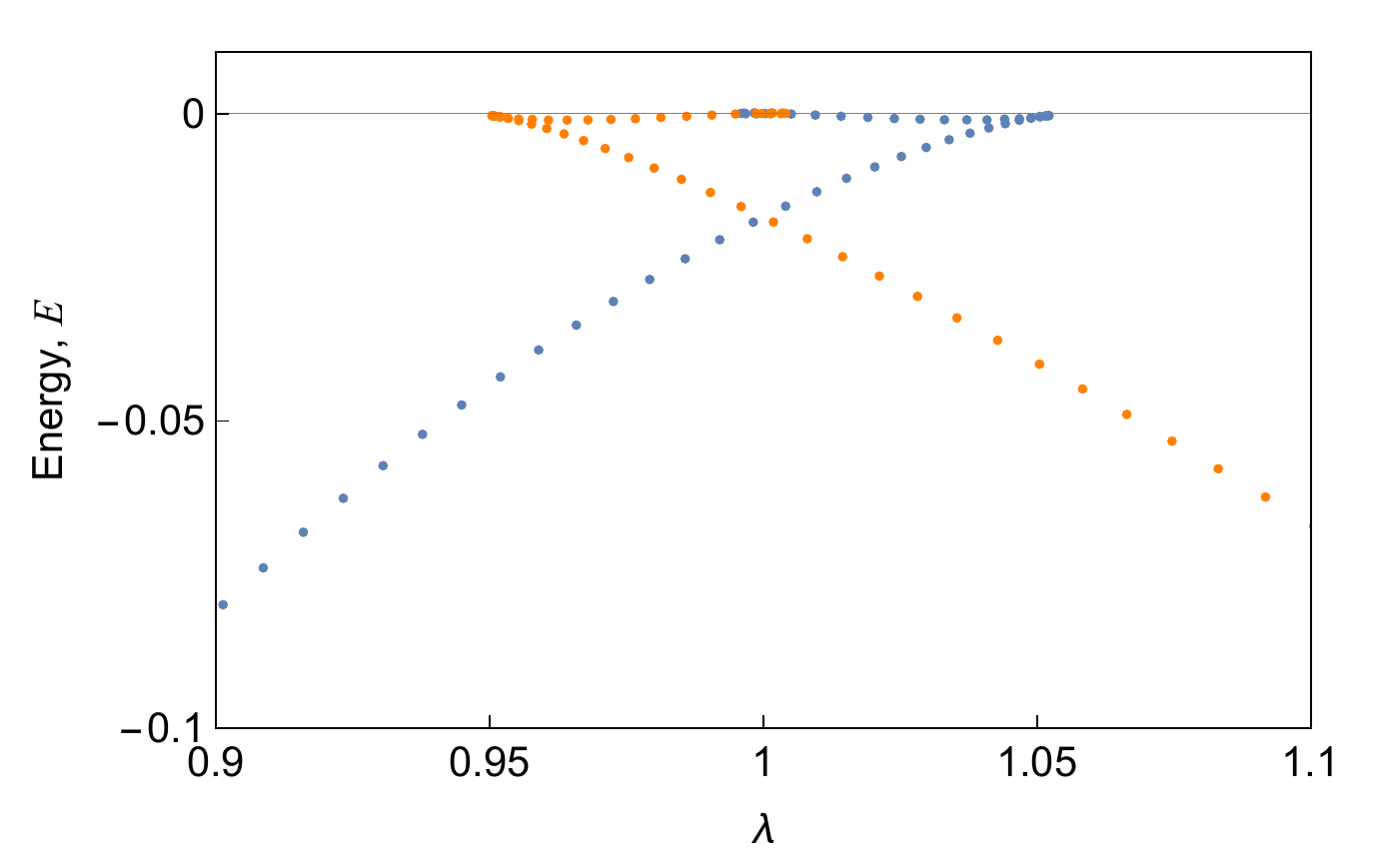}
    \end{subfigure}
    \caption{Energy for the two families of ground states in AdS${}_6$. The blue curve corresponds to solutions with the first 2-sphere pinching off and the orange curve corresponds to the second 2-sphere pinching off. The two types of ground states switch dominance at $\l=1$. The oscillatory behaviour of the solutions with smaller $r_0$ is reflected in an oscillation in the energy plot, but they have higher energy and do not dominate.}%
    \label{fig:AdS6_2SphereAction}
\end{figure}
The other family of solutions where the second sphere pinches off can be obtained by relabelling and conformal rescaling, with the effect of $\l\to 1/ \l$.

Fig.~\ref{fig:AdS6_2SphereAction} shows the energy for these non-black hole solutions. The two curves are related by relabelling: $E(\l)=\tilde E\left(1/\l\right)/\l$, where $\tilde E(\l)$ is the lowest energy  solution where the first sphere is contractible and $E(\l)$ is when the second sphere is contractible.\footnote{Due to the absence of a conformal anomaly in odd dimensions, the energy does not receive extra correction.}

\begin{figure}[ht]
    \centering
    \includegraphics[width=10cm]{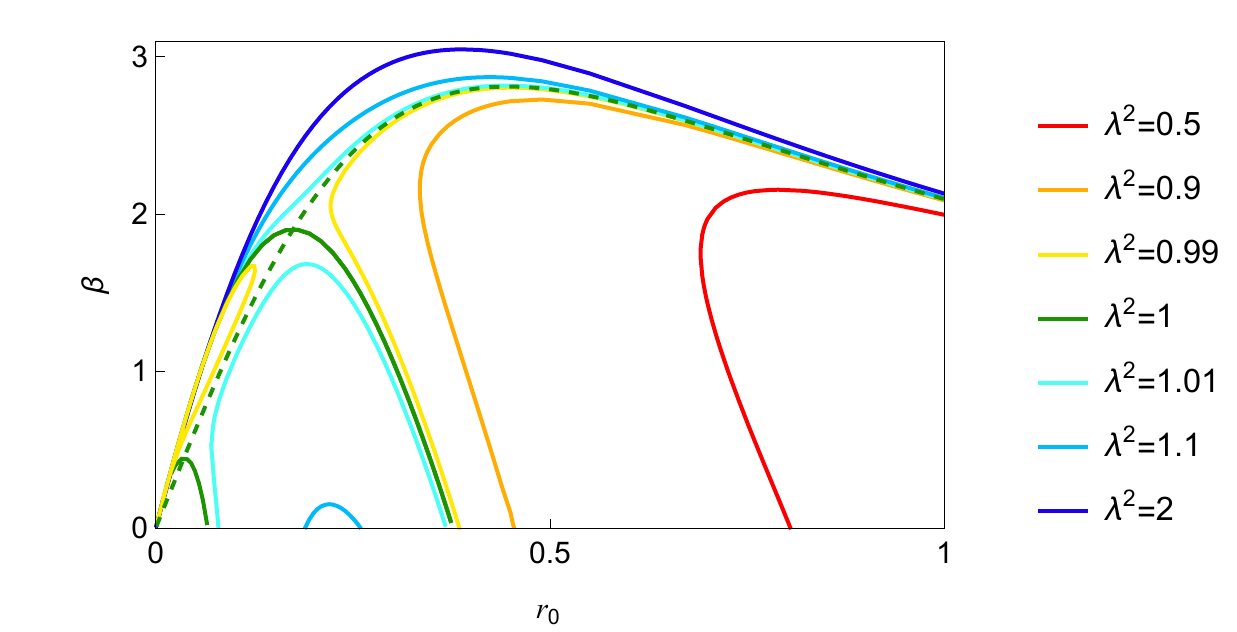}
    \caption{Inverse temperature versus $r_0$ for various $\l$. The dashed curve corresponds to the analytic solution. Several branches of solutions with the same $\l$ are  visible. }
    \label{fig:AdS6_Tvsr0}
\end{figure}

\begin{figure}[ht]
    \centering
    \includegraphics[width=8cm]{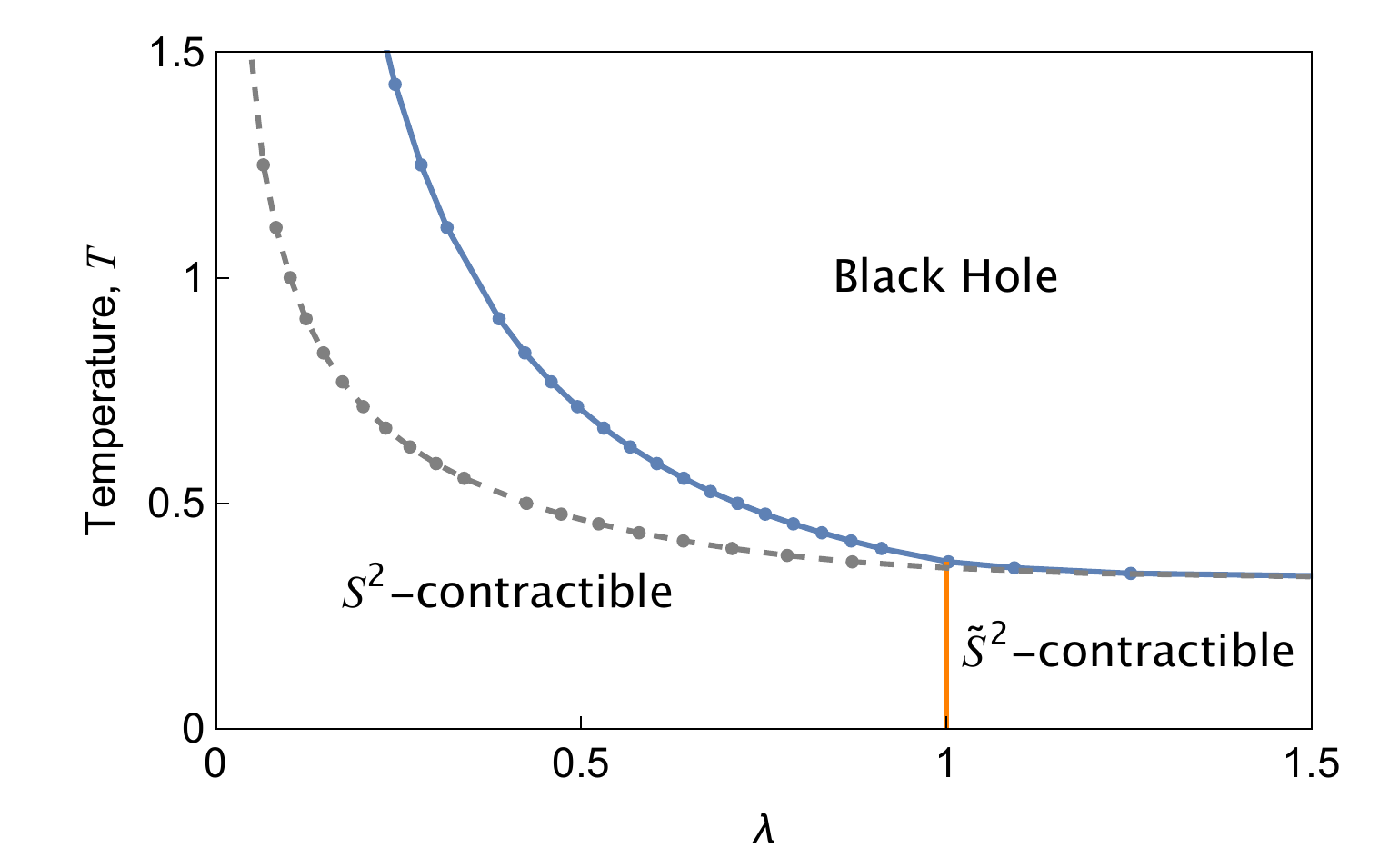}
    \caption{Phase diagram for AdS${}_6$ in the thermal ensemble with $S^1\times S^2\times S^2$ boundary conditions. The blue curve demarcates a transition between black holes and non-black holes, and the orange line separates two ground states, where the solution with the first $S^2$ contracting dominates at small $\lambda$ and the second $S^2$ at large $\l$. The gray dashed curve is the minimum temperature curve for a black hole solution to exist.}
    \label{fig:AdS6_Phase}
\end{figure}
\begin{figure}[ht]
    \centering
    \includegraphics[width=8cm]{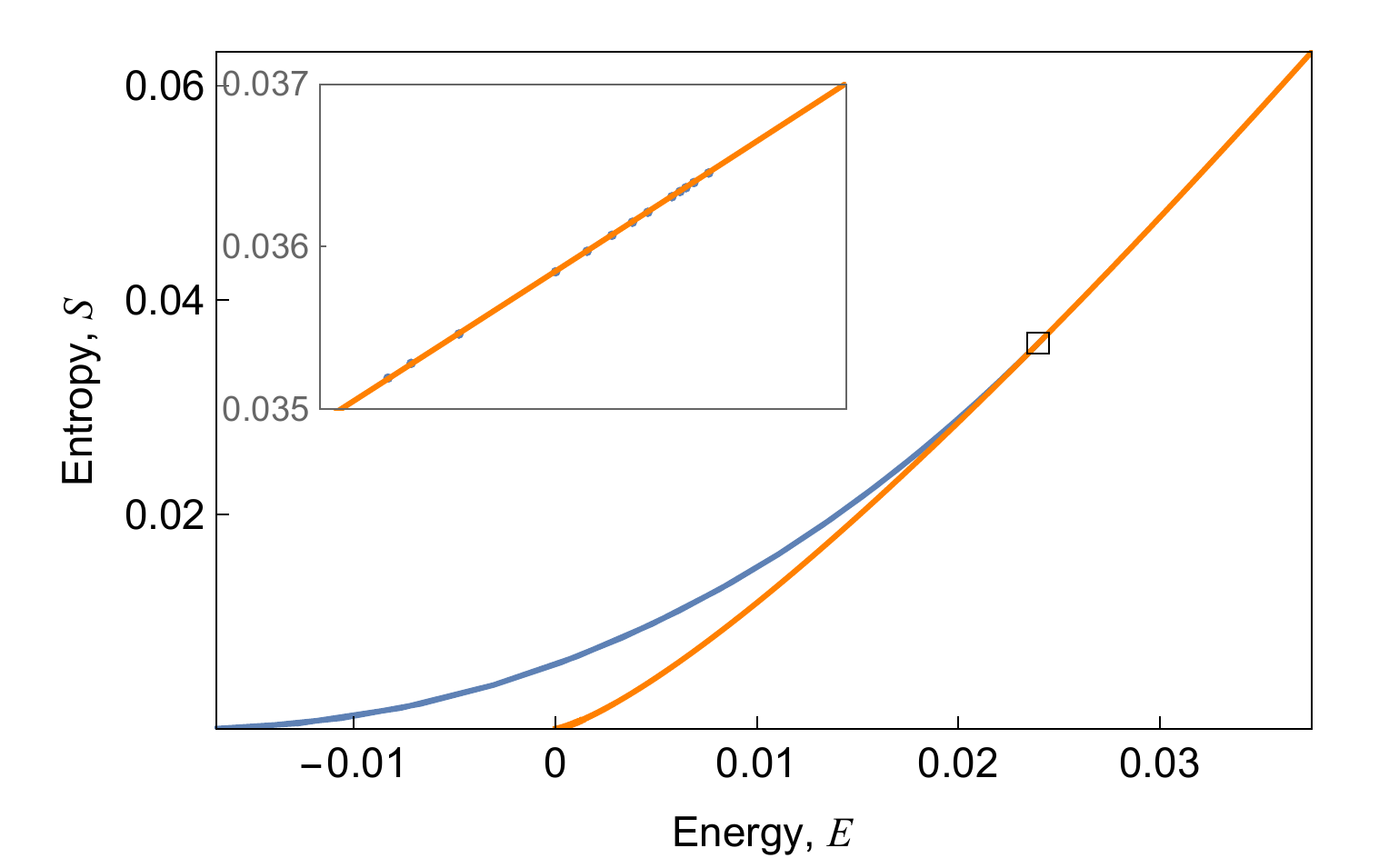}
    \caption{Entropy-energy plot for the analytic black hole solutions (shown in orange) and the first numerical branch of black holes (shown in blue) at the critical $\l=1$. }
    \label{fig:AdS6_micro}
\end{figure}

For black holes, numerical solutions were found in \cite{Hickling:2016mzp} where the two-parameter family of solutions were labelled by the size of each of the spheres at the horizon. Here we focus on their dependence on $\l$ and study the number of one-parameter families of solutions at any given $\l$. Just as in the main text, we find a growing number of solutions near $\l_c$. Again, there is an analytic branch at this special ratio, given by
\begin{equation}
    ds^2 = - \mathcal{A}(r) dt^2 +\frac{dr^2}{\mathcal{A}(r)} + r^2 \left [d\Omega_2^2   + d\tilde\Omega_2^2\right],
    \end{equation}
where 
\begin{equation}
   \mathcal{A}(r)=r^2+\frac{1}{3}-\frac{r_0^3+3r_0^5}{3r^3}.
\end{equation}
Fig.~\ref{fig:AdS6_Tvsr0} shows a few branches of solutions for several $\l$.

In the canonical ensemble, the phase diagram is similar to that of AdS$_7$ and shown in Fig.~\ref{fig:AdS6_Phase}. We can see that there are again three phases, corresponding to three types of bulk configurations. The line separating the black hole from the rest is a one-parameter generalization of the Hawking-Page phase transition. The critical temperature depends on $\l$. The vertical line separating the two types of non-black hole solutions is independent of the temperature. At zero temperature, this is a quantum phase transition.

In the microcanonical ensemble, we look at the entropy versus energy plot Fig.~\ref{fig:AdS6_micro}. There is again clear evidence for a phase transition, but the numerics are not accurate enough to determine if the two families of solutions cross or become tangent. So it is not clear if the transition is first or second order.

\bibliographystyle{JHEP}
\bibliography{library}

\end{document}